\documentclass[fleqn,usenatbib]{mnras}

\usepackage{times}

\usepackage[T1]{fontenc}
\usepackage{xcolor}
\usepackage{ae,aecompl}
\usepackage{graphicx}
\usepackage{amsmath}
\usepackage{amssymb}

\title[Cosmological SPH simulations]{An improved SPH scheme for cosmological simulations}
\author[A. M. Beck et al.]{A. M. Beck$^{1,2}$\thanks{E-mail: abeck@usm.lmu.de (AMB)}, G. Murante$^{3}$, A. Arth$^{1,4}$, R.-S. Remus$^{1}$, A. F. Teklu$^{1,5}$, J. M. F. Donnert$^{6,7}$\thanks{E-mail: donnert@ira.inaf.it (JMFD)}, \newauthor S. Planelles$^{8,3}$, M. C. Beck$^{9}$, P. F\"{o}rster$^{1}$, M. Imgrund$^{1}$, K. Dolag$^{1,2}$ and S. Borgani$^{3,8,10}$\\
  $^{1}$University Observatory Munich, Scheinerstr. 1, D-81679 Munich, Germany\\
  $^{2}$Max Planck Institute for Astrophysics, Karl-Schwarzschild-Str. 1, D-85741 Garching, Germany\\
  $^{3}$INAF - Osservatorio Astronomico di Trieste, via Tiepolo 11, I-34131 Trieste, Italy\\
  $^{4}$Max Planck Institute for Extraterrestrial Physics, Giessenbachstr. 1, D-85748 Garching, Germany\\
  $^{5}$Excellence Cluster Universe, Boltzmannstr. 2, D-85748 Garching, Germany\\
  $^{6}$INAF - Instituto di Radioastronomia, via P. Gobetti 101, I-40129 Bologna, Italy\\
  $^{7}$Leiden Observatory, Leiden University, PO Box 9513, NL-2300 Leiden, the Netherlands\\
  $^{8}$Department of Physics, University of Trieste, via Tiepolo 11, I-34131 Trieste, Italy\\
  $^{9}$Department of Physics, University of Konstanz, Universit\"{a}tsstr. 10, D-78457, Germany\\
  $^{10}$INFN - Sezione di Trieste, via Valerio 2, I-34127 Trieste, Italy}

\date{Accepted XXX. Received YYY; in original form ZZZ}

\pubyear{2015}

\begin{document}

\label{firstpage}
\pagerange{\pageref{firstpage}--\pageref{lastpage}}

\maketitle

%#########################################################################################################
%########################### Abstract ####################################################################
%#########################################################################################################

\begin{abstract}
We present an implementation of smoothed particle hydrodynamics (SPH) with improved accuracy for simulations of galaxies and the large-scale structure.
In particular, we implement and test a vast majority of SPH improvement in the developer version of GADGET-3.
We use the Wendland kernel functions, a particle wake-up time-step limiting mechanism and a time-dependent scheme for artificial viscosity including high-order gradient computation and shear flow limiter.
Additionally, we include a novel prescription for time-dependent artificial conduction, which corrects for gravitationally induced pressure gradients and improves the SPH performance in capturing the development of gas-dynamical instabilities.

We extensively test our new implementation in a wide range of hydrodynamical standard tests including weak and strong shocks as well as shear flows, turbulent spectra, gas mixing, hydrostatic equilibria and self-gravitating gas clouds.
We jointly employ all modifications; however, when necessary we study the performance of individual code modules.
We approximate hydrodynamical states more accurately and with significantly less noise than standard GADGET-SPH.
Furthermore, the new implementation promotes the mixing of entropy between different fluid phases, also within cosmological simulations.

Finally, we study the performance of the hydrodynamical solver in the context of radiative galaxy formation and non-radiative galaxy cluster formation.
We find galactic disks to be colder and more extended and galaxy clusters showing entropy cores instead of steadily declining entropy profiles.
In summary, we demonstrate that our improved SPH implementation overcomes most of the undesirable limitations of standard GADGET-SPH, thus becoming the core of an efficient code for large cosmological simulations.
\end{abstract}

%#########################################################################################################
%########################### Keywords ####################################################################
%#########################################################################################################


%#########################################################################################################
%########################### Introduction ################################################################
%#########################################################################################################

\section{Introduction}

Smoothed particle hydrodynamics (SPH) is a commonly employed numerical method in astrophysics.
It solves the fluid equations \citep{landau59} in a Lagrangian mass-discretised fashion, which ensures Galilean invariance and conservation of mass, momentum, angular momentum, energy and entropy.
It was pioneered by \cite{gingold77} and \cite{lucy77} and has since then become one of the cornerstones of computational astrophysics.
The discretisation of mass automatically adapts spatial resolution by removing the constraint of handling geometry explicitly.
It also easily couples to N-Body schemes for calculation of gravitational forces \citep{hernquist89}.
An excessive amount of papers and literature about SPH has been produced over the past decades.
We point out the latest reviews by \cite{rosswog09}, \cite{springel10b}, \cite{monaghan12} and \cite{price12a} for the basic concepts and e.g. \cite{ritchie01} for an extension to multi-phase fluids and \cite{rosswog14} for a special-relativistic adaption.
As every numerical method, SPH comes with its own set of benefits and pitfalls, which we address in this paper.

The inability of traditional SPH methods to treat contact discontinuities and to mix different fluid phases is a long-standing problem \citep{agertz07,wadsley08}.
It leads to a completely numerical spurious surface tension at the discontinuities preventing particle movements.
Consequently, it results in a failure of these formulations of SPH to resolve fluid instabilities such as the Kelvin-Helmholtz or Rayleigh-Taylor instabilities \citep[see e.g.][]{junk10,valcke10,mcnally12,puri13}.
In applications to cosmic structure formation it causes entropy profiles to diverge towards the centres of dark matter haloes, at variance with Eulerian codes that predict entropy plateaus to build up \citep{frenk99,wadsley08,planelles09,vazza11,power14,biffi15,rasia15}.
This difference is due to the lack of mixing in simple SPH, which makes low-entropy gas in merging substructures sink toward the centre of the main structure.
Many modifications have been proposed to overcome this problem.
For example, \cite{wadsley08} propose a mixing solution, which resolves the differences in the entropy profiles of dark matter haloes between Eulerian and SPH codes \citep{frenk99}.
Further cosmological applications have been performed by \cite{shen10}.
Firstly, the equation of motion (EoM) can be re-formulated from a standard 'density' approach into a 'pressure' based approach \citep{saitoh13,hopkins13}.
While the 'pressure' formulation correctly treats contact discontinuities, it leads to increased noise at strong shocks.
Secondly, considerable effort has been made to unite grid-based solvers for the fluid equations with the Lagrangian nature of SPH.
Eulerian Godunov methods \citep[see e.g.][]{cha10,springel10a,murante11} and their coupling to Lagrangian methods is a promising alternative.
Connecting a Lagrangian moving-mesh with grid-based solvers \citep{springel10a} or mesh-free approaches \citep{hopkins15a,hopkins15b} represent more advanced approaches.
Thirdly, artificially modelled conduction (AC) of internal energy can be employed to overcome the mixing problem.
Most modern SPH codes include AC of some sort to diffuse entropy \citep{read12} or energy \citep{price08} across particles.
The use of AC has to be taken carefully and it is only desirable at contact discontinuities in traditional 'density' SPH and at shocks in modern 'pressure' SPH.
The application of AC in other regions can have catastrophic impact on the fluid dynamics and can smear out gravitationally established pressure gradients, thus leading to totally numerically induced transport of internal energy \citep{valdarnini12}.

Next, traditional SPH has difficulties treating subsonic turbulence as it experiences a high effective viscosity, which limits the inertial range \citep[see e.g.][]{bauer12}.
Thus, traditional SPH cannot achieve high Reynolds' numbers compared to, for exmaple, Eulerian methods.
This high effective viscosity is a function of resolution, but no general solution has yet been proposed to resolve this issue in general.
For the correct capturing of shocks numerically motivated artificial viscosity (AV) is commonly employed.
It smooths the particle velocity distribution and gives order to the fluid sampling.
However, AV is only desired at the shock and the fluid should be inviscid otherwise.
Too much AV smears out physical motions and damps subsonic and turbulent motions in isolated tests \citep{bauer12} as well as in cosmological simulations \citep{dolag05}.
Therefore, several different implementations of AV reduction are proposed \citep{morris97,cullen10}.
They are all based around a proper shock detection method and a time-dependent viscosity decay scheme.
Application of such advanced schemes give better results in the description of fluid motions \citep{dolag05,price12b}.

Finally, it might seem easy to simply reduce the quantitative errors by increasing the number of neighbours, which contribute to the local density and force estimators.
However, the standard  weighting functions of SPH respond differently to an increase of smoothing neighbours and can possibly become unstable to the pairing instability \citep{schuessler81,price12a}.
Therefore, recently, alternative kernel functions immune against this instability are proposed for better fluid sampling and convergence \citep{read10,dehnen12}.
The advantage of flexible geometry of SPH comes with difficulties in creating well-defined initial conditions or sampling analytical profiles, where we use either glass set-ups \citep{white96} or Weighted Voronoi-Tesselations \citep{diehl12}.

To overcome the named disadvantages we implement a large set of improvements for SPH into the developer version of the cosmological N-Body / SPH simulation code GADGET-3 \citep{springel01,springel05}.
We include a time-step limiter for strong shocks, a time-dependent viscosity scheme for subsonic turbulence, a high-order gradient estimator and shear flow limiter for shearing motions, an improved kernel function for convergence and a time-dependent artificial conduction scheme to promote fluid mixing.
We discuss the accuracy and the performance of our new scheme in hydrodynamical standard test problems, within quiet and violent environments as well as in Idealized simulations of galaxy and galaxy cluster formation, in which our new scheme is applied to reasonable astrophysical problems.

The paper is organised as follows.
The improved implementation of hydrodynamics is presented in Section 2.
In Section 3 we validate our SPH algorithms in a set of hydrodynamical standard tests and we proceed to standard tests with gravity in Section 4.
We continue in Section 5 with Idealized applications to the evolution of an isolated disk galaxy and a forming galaxy cluster.
We summarise our developments and code performance in Section 6.

%#########################################################################################################
%########################### Hydrodynamics ###############################################################
%#########################################################################################################

\section{A new SPH implementation}

We start with a presentation of the main equations corresponding to a `standard' and our `new' formalism of GADGET-SPH.
The formalism of SPH is already well described by a large number of reviews \citep[see e.g.][]{price12a}.
We refer to the `standard' version of SPH as the implementation within the GADGET-3 code without our modifications.
We point out our modifications and discuss the kernel function, the EoM, the particle wake-up scheme and the time-dependent AV and AC.

\subsection{Original code platform}

\begin{table*}
\begin{center}
  \begin{tabular}{@{}llll}
    \hline\hline
    & Standard & New & \\\hline\hline
    Density estimator & Traditional & Bias-corrected & \cite{dehnen12} \\
    Kernel function & Cubic spline & Wendland C$^{4}$ & \cite{dehnen12} \\
    Neighbours (3D) & 64 & 200 & \cite{dehnen12} \\\hline
    Equation of motion & Density-Entropy & Density-Entropy & \cite{springel02} \\
    Grad-$h$ terms & Yes & Yes & \cite{springel02} \\\hline
    Velocity gradients & Low-order & High-order & \cite{price12a,hu14} \\
    Artificial viscosity & Standard (constant) & Adaptive (locally) & \cite{dolag05,cullen10} \\
    Balsara limiter & Low-order & High-order & \cite{balsara95,cullen10,price12a} \\\hline
    Artificial conduction & No & Adaptive (locally) & \cite{wadsley08,price08} \\
    Hydrostatic correction & No & Adaptive (locally) & \cite{price08,valdarnini12} \\\hline
    Particle wake-up & No & yes (f$_{w}=3$) & \cite{saitoh09,pakmor10,pakmor12}\\\hline\hline
  \end{tabular}
  \caption{Comparison of the `standard' (column 2) and `new' (column 3) SPH implementations in the GADGET code.
  Furthermore, we give some references (column 4) for extended descriptions and discussions.}
  \label{tab:comp}
\end{center}
\end{table*}

We implement our SPH modifications into the developer version of the cosmological N-Body/SPH code GADGET-3 \citep{springel01,springel05}.
We evolve entropy as the thermodynamical variable \citep{springel02} and use the prescriptions for radiative cooling, supernova feedback and star formation following \cite{springel03}.
In the following sections we compare two different SPH schemes (see also Table \ref{tab:comp}), which are distinguished as follows.
The `standard' implementation corresponds to the developer version of GADGET-3 without our modifications \citep{springel05}.
The `new' implementation includes all the SPH improvements presented in this section.
In principle, we always use the entire new scheme and we also employ the same set of all numerical parameters throughout our entire simulation test suite.
Thus, unless otherwise stated, we do not tune individual standard tests or astrophysical applications.
However, if necessary we sometimes switch off some of the modifications to analyse their individual and isolated impact on several of the test problems.

\subsection{Kernel functions and density estimate}

Foremost, there is the question in a Lagrangian method how to derive fluid field quantities from a given set of point masses.
In particular, the estimation of the gas density is crucial as many further equations rely on it.
We employ the standard estimator of SPH and calculate the density $\rho(\mathbf{x}_{i})=\rho_{i}$ of an individual particle $i$ at the position $\mathbf{x}_{i}$ by summing the contributions of $N_i$ neighbouring particles $j$ within a smoothing radius $h(\mathbf{x}_{i})$ at a distance $x_{ij}$ in a mass-weighted ($m_j$) and distance-weighted ($W_{ij}(x_{ij},h_{i})$) fashion

\begin{equation}\rho_{i} = \sum_{j}m_{j}W_{ij}(x_{ij},h_{i}).\label{eq:de}\end{equation}
Simultaneously, the smoothing length $h_i$ is a function of density

\begin{equation}h(\mathbf{x}_{i}) = \eta\left(\frac{m_i}{\rho_i}\right)^{1/3},\label{eq:he}\end{equation}
where $\eta$ defines the ratio of smoothing length to the mean distance between particles.
Eqs. (\ref{eq:de}) and (\ref{eq:he}) roughly ensure constant mass resolution throughout the simulation and have to be solved in parallel.
This mimics the evolution of spheres of the same mass $4/3\pi{}h_i^3\rho_i = N_{i}m_{i}$ but with varying number of neighbours.
The number of neighbours varies across space and time with an increase or decrease of smoothing length and local quality of fluid sampling by the point masses.
The weighting function is commonly chosen to decrease monotonically with distance, yield smooth derivatives, is symmetric with respect to $x_{ij}=x_{ji}$ and has a flat central portion.
A historical choice \citep{monaghan85} of kernel function 

\begin{equation}W_{ij}(x_{ij},h_{i})=w(q)/h_{i}^{3},\end{equation}
is the cubic B-spline function with $q=x_{ij}/h_{i}$ and 

\begin{equation}
   w(q)=\frac{8}{\pi}\left\{\begin{array}{ll}
      1 - 6 q^2 + 6 q^3 \;\;& 0 \le q \le \frac{1}{2} \\
      2 \left(1 - q\right)^3                              & \frac{1}{2} \le q \le 1 \\
      0                                                             & 1 \le q \\
   \end{array} \right.,
\end{equation}
which we commonly employ with a choice of 64 neighbours in three dimensions.
However, this traditional kernel function is subject to the pairing (or clumping) instability when the number of neighbours is too large \citep[see][]{price12a}.
An alternative choice to achieve better numerical convergence is necessary and an entire new family of kernels is needed.
In a kernel stability analysis \cite{dehnen12} show that the Wendland kernel functions are a much better choice.
We choose the Wendland $C^{4}$ (WC4) kernel with 200 neighbours in three dimensions as our smoothing function without the pairing instability problem.
The functional form of the $C^{4}$ is given by

\begin{equation}w(q)=\frac{495}{32\pi}(1-q)^{6}(1+6q+\frac{35}{3}q^{2}).\end{equation}
For values of $q>1$ it is set to $w(q)=0$.
The Wendland functions require similar computational effort as the cubic spline kernel but nevertheless, the total computational time increases due to larger number of neighbours.
Therefore, we do not employ the higher-order $C^{6}$ functions because of the required 295 neighbours.
In summary, the total computational cost of the density and hydrodynamical force calculation increases by a factor of about two in comparison to a cubic spline with 64 neighbours.
However, a better estimate of the kernel will result in a more accurate density estimate and improved gradient estimators.
These estimators are the cornerstones of the SPH formalism and determine the accuracy and convergence rate in all our test simulations.

\subsection{Equation of motion}

The EoM for a system of point masses are derived \citep[see e.g.][]{price12a} from a discretised version of the fluid Lagrangian

\begin{equation}L=\sum_i{m_i\left[\frac{1}{2}v^2_i-u_i\right]},\end{equation}
where $v$ denotes velocities and $u$ internal energy of individual particles.
The Lagrangian nature of SPH, when complemented with a symplectic time integration scheme, automatically conserves mass, momentum, angular momentum, energy and entropy.
We use the standard kick-drift-kick Leapfrog time integration of GADGET \citep{springel05}.
The EoM then follows from the principle of least action, where the spatial derivative of internal energy comes (if constant entropy is assumed) from the first law of thermodynamics $dU=-PdV$.
We choose a volume element depending on density ($V=m/\rho$) and an adiabatic equation of state for the pressure $P=A\rho^{\gamma}$, which is defined individually for every particle.
We integrate entropy $A$ as the thermodynamical variable of choice and thus employ what is commonly called 'density-entropy' SPH.

The EoM in the 'density-entropy' \citep[for a derivation see][]{springel02} for the hydrodynamical force part of an individual particle reads

\begin{equation}\left.\frac{d\mathbf{v}_{i}}{dt}\right|_{\mathrm{hyd}} = -\sum_{j}m_{j}\left[f_{i}^\mathrm{co}\frac{P_{i}}{\rho_{i}^2}\mathbf{\nabla}_{i}W_{ij}(h_{i})+f_{j}^\mathrm{co}\frac{P_{j}}{\rho_{j}^2}\mathbf{\nabla}_{i}W_{ij}(h_{j})\right],\label{eq:eom}\end{equation}
where the factor $f^\mathrm{co}$ is a correction factor, which accounts for the mutual co-dependence of smoothing length $h(\rho)$ and density $\rho(h)$ and their corresponding derivatives.
Its functional form is given by

\begin{equation}f_{i}^\mathrm{co}=\left[1+\frac{h_{i}}{3\rho_{i}}\frac{\partial\rho_{i}}{\partial{}h_{i}}\right]^{-1}.\end{equation}
Eq. (\ref{eq:eom}) leads to a non-vanishing force at contact discontinuities even when pressure is constant.
This is the artificial 'surface tension' of SPH, which suppresses particle movement across contact discontinuities.
In the following sections, we present our equations in notation of internal energy $u$, which is related to the entropic function $A=(\gamma-1)u/\rho^{\gamma-1}$.

\subsection{Artificial viscosity}\label{sec:visc}

\subsubsection{Smoothing of jumps}

By construction, SPH solves the ideal Euler equation and no dissipative terms are included but those are necessary to describe discontinuities correctly.
In highly dynamical regions (e.g. in shocks) fast particles commonly penetrate into regions of resting particles causing unwanted particle disorder and oscillations in the sampling of the fluid.
However, SPH already contains an intrinsic remeshing force but to re-establish particle order and capture shocks properly an additional dissipative term in velocity is needed.
This AV aims to remove post-shock oscillations and noise and helps to smooth the velocity field \citep[see][]{monaghan83}.
We include AV in an energy conserving way with a contribution to the EoM of the form
\begin{equation}\left.\frac{d\mathbf{v}_{i}}{dt}\right|_{\mathrm{visc}} = \frac{1}{2}\sum_{j}\frac{m_{j}}{\rho_{ij}}\left(\mathbf{v}_{j}-\mathbf{v}_{i}\right)\alpha^{v}_{ij}f_{ij}^\mathrm{shear}v^\mathrm{sig,v}_{ij}\overline{F}_{ij},\end{equation}
and with a contribution to the energy equation of the form

\begin{equation}\left.\frac{du_{i}}{dt}\right|_{\mathrm{visc}} = -\frac{1}{2}\sum_{j}\frac{m_{j}}{\rho_{ij}}\left(\mathbf{v}_{j}-\mathbf{v}_{i}\right)^{2}\alpha^{v}_{ij}f_{ij}^\mathrm{shear}v^\mathrm{sig,v}_{ij}\overline{F}_{ij},\end{equation}
where the symmetrised variables represent $\rho_{ij}=(\rho_{i}+\rho_{j})/2$ for the density, $\alpha^{v}_{ij}=(\alpha^{v}_{i}+\alpha^{v}_{j})/2$ as a numerical coefficient to include AV (see below) and $f_{ij}^\mathrm{shear}=(f^\mathrm{shear}_{i}+f^\mathrm{shear}_{j})/2$ as the \cite{balsara95} shear flow limiter (see Section 2.3.2 below), which aims to ensure the application of AV only in strong shocks (high velocity divergence) and not in rotating or shearing flows (high velocity curl).
Furthermore, in the above equation $\overline{F}_{ij}=(F_{ij}(h_i)+F_{ij}(h_j))/2$ is the symmetrised scalar part of the kernel gradient terms $\mathbf{\nabla}_{i}W_{ij}(h_i)=F_{ij}\mathbf{r}_{ij}/r_{ij}$, which are used to linearly interpolate the second-order Laplacian derivative in the velocity field diffusion equation.
The pairwise signal velocity $v^\mathrm{sig,v}_{ij}$ \citep[first introduced by][and already used in GADGET-2]{monaghan97} determines the strength of AV and directly includes a quantitative measure of particle disorder

\begin{equation}v_{ij}^\mathrm{sig,v}=c^{s}_{i}+c^{s}_{j}-\beta\mu_{ij}\label{equ:signal},\end{equation}
where $c^{s}$ is the sound speed of the particles and $\mu_{ij}=\mathbf{v}_{ij}\cdot\mathbf{x}_{ij}/x_{ij}$ with a commonly chosen pre-factor of $\beta=3$.
AV is only applied between approaching pairs of particles (i.e. $\mu_{ij}<0$) and otherwise switched off.
The local signal velocity $v_{i}^\mathrm{sig}$ (also used by the time-step criterion, see Section $\ref{sec:wakeup}$) represents the maximum value of $v_{ij}^\mathrm{sig,v}$ between all particle pairs $ij$ within the entire smoothing sphere of particle $i$.

The calculation of the viscosity coefficient $\alpha^{v}_{i}$ is based on an approach developed by \cite{cullen10} but modified for more efficient computation as follows.
The presence of a shock is indicated via computation of velocity divergence contributions across the entire smoothing length by

\begin{equation}R_i=\frac{1}{\rho_{i}}\sum_{j}{\mathrm{sign}(\mathbf{\nabla}\cdot\mathbf{v})_{j}m_{j}W_{ij}},\end{equation}
where a shock corresponds to $R_{i}\approx{}-1$.
In principle, an accurate calculation of $R_i$ for every particle requires the previous computation of $(\mathbf{\nabla}\cdot\mathbf{v})_{i}$ for every particle.
Therefore, an extra SPH summation loop added between the calculation of density (where velocity divergence can also be calculated) and hydro forces would be necessary.
For computational reasons we use the velocity divergence calculated in the previous time-step.
Furthermore, a convergent flow is also indicated by a high velocity divergence but that condition does not distinguish between pre-shock and post-shock regions.
Therefore, we employ the time derivative of velocity divergence to determine a directional shock indicator

\begin{equation}A_i=\xi_{i}\mathrm{max}(0,-(\dot{\mathbf{\nabla}}\cdot\mathbf{v})_{i}),\label{eq:ax}\end{equation}
which is able to distinguish between pre-shock and post-shock regions.
We calculate $(\dot{\mathbf{\nabla}}\cdot\mathbf{v})_{i}$ via interpolation between the current and the previous time-step \citep[as suggested by][]{cullen10} in the time interval $\Delta{}t_{i}$.

Subsequently, we use the shock indicator $R_i$ to determine the ratio $\xi_i$ of strength of shock and strength of shear in quadratic form via

\begin{equation}\xi_i = \frac{|2(1-R_i)^{4}(\mathbf{\nabla}\cdot\mathbf{v})_{i}|^{2}}{|2(1-R_i)^{4}(\mathbf{\nabla}\cdot\mathbf{v})_{i}|^{2}+|\mathbf{\nabla}\times{}\mathbf{v}|_{i}^2},\end{equation}
which is proposed by \cite{cullen10} as an additional limiting factor for AV in Eq. (\ref{eq:ax}) and was experimentally determined.
Now, for every particle we can define and set the target value $\alpha_{i}^{\mathrm{loc},{v}}$ of AV with the help of the directional shock indicator to

\begin{equation}\alpha_{i}^{\mathrm{loc},{v}}=\alpha_\mathrm{max}\frac{h^{2}_{i}A_{i}}{h^{2}_{i}A_{i}+(v_{i}^\mathrm{sig})^{2}}.\end{equation}
In the case, where the viscosity coefficient $\alpha_{i}^{v}$ is smaller than $\alpha_{i}^{\mathrm{loc},{v}}$, we set the coefficient to $\alpha_{i}^{\mathrm{loc},{v}}$.
Otherwise, we let it decay with time according to

\begin{equation}\dot{\alpha}^{v}_{i}=(\alpha_{i}^{\mathrm{loc},{v}}-{\alpha}^{v}_{i})\frac{v_{i}^\mathrm{sig}}{\ell{}h_{i}},\end{equation}
which we integrate in time together with the hydrodynamical quantities.
The constant $\ell$ specifies the decay length and in our test problems we find a numerical value of $\ell={}4.0$ to give reasonable results.

\subsubsection{Gradient estimators}

We use the \cite{balsara95} form of the shear viscosity limiter

\begin{equation}f_{i}^\mathrm{shear} =\frac{|\mathbf{\nabla}\cdot\mathbf{v}|_i}{|\mathbf{\nabla}\cdot\mathbf{v}|_i+ |\mathbf{\nabla}\times\mathbf{v}|_i+\sigma_{i}},\end{equation}
with $\sigma_{i}=0.0001c^{s}_{i}/h_{i}$ for numerical stability reasons.
At this point, the question arises how to calculate the divergence and vorticity.
The common curl estimator of SPH reads

\begin{equation}(\mathbf{\nabla}\times{}\mathbf{v})_{i}=-\frac{1}{\rho_{i}}\sum_{j}m_{j}\left(\mathbf{v}_{j}-\mathbf{v}_{i}\right)\times\mathbf{\nabla}_iW_{ij},\label{eq:div}\end{equation}
which takes the lowest order error term into account.
Since higher-order error terms are neglected, this formation performs very poorly in the regime of strong shear flows.
Therefore, we resort to a higher-order calculation scheme of the velocity gradient matrix \citep[see similar approaches by][]{cullen10,price12a,hu14}.
We follow the approach presented in \cite{price12a} for the computation of the gradient matrix.
We expand $\mathbf{v}_{j}$ for every vector component $k$ in a Taylor-series around $i$ with
\begin{equation}v^k_{j}=v^k_{i}+(\partial_\delta{}v^k_{i})(\mathbf{x}_{j}-\mathbf{x}_{i})^\delta+O(h^{2}).\label{eq:tay}\end{equation}
Inserting Eq. (\ref{eq:tay}) into Eq. (\ref{eq:div}) yields an easy solution for the linear term $\partial_\delta{}v^k_{i}$ and the velocity gradient matrix by solving the following system of equations:

\begin{equation}\chi^{\alpha\beta}=\sum_{j}m_{j}(\mathbf{x}_{j}-\mathbf{x}_{i})^{\alpha}\mathbf{\nabla}_i^{\beta}W_{ij},\end{equation}
\begin{equation}\chi^{\alpha\beta}\frac{\partial{}v^{k}_{i}}{\partial\mathbf{x}^\alpha}=\sum_{j}m_{j}(\mathbf{v}_{j}-\mathbf{v}_{i})^{k}\mathbf{\nabla}_i^{\beta}W_{ij},\end{equation}
which requires a matrix inversion for $\chi^{\alpha\beta}$.
Conveniently, the estimator is also independent of density and thus, can be calculated in the same computational loop along with densities.
Subsequently, the updated estimates of velocity divergence and curl are calculated directly from the full velocity gradient matrix via

\begin{equation}(\mathbf{\nabla}\cdot{}\mathbf{v})_{i}=\frac{\partial{}v_{i}^{\alpha}}{\partial\mathbf{x}^\alpha},\end{equation}
\begin{equation}(\mathbf{\nabla}\times{}\mathbf{v})_{i}^{\delta}=\epsilon_{\alpha\beta\delta}\frac{\partial{}v_{i}^{\beta}}{\partial\mathbf{x}^\alpha}.\label{eq:high_curl}\end{equation}
In our test problems we find the low-order estimator of velocity divergence to give already satisfying results \citep[see also appendix A2 in][]{schaye15}.
In contrast, the low-order estimator of velocity curl performs very poorly and we obtain significantly improved results with the high-order curl estimator of Eq. (\ref{eq:high_curl}).
The high-order estimators are not restricted to the AV scheme but they also enter various other modules of the code, where their precise calculation is required.
For example, this additionally greatly improves the approximation of fluid vorticity written into the simulation snapshots.

\subsection{Artificial conductivity with gravity correction}

\subsubsection{Smoothing of jumps}

We move on to address the mixing problem in SPH by introducing a kernel-scale exchange term for internal energy transport.
We include AC for purely numerical reasons to treat discontinuities in the internal energy (similar to the capturing of velocity jumps by AV), which arise from our 'density-entropy' formulation of SPH.
We note that a 'pressure-entropy' formulation of the EoM is also able to address the mixing problem but it also requires the presence of AC in order to smooth noise in internal energy behind shocks \citep{hopkins13,hu14}.
Thus, in either flavor of SPH the inclusion of AC is recommended and many different formulations of the AC diffusion equation have been investigated so far.
Although their precise details vary across the literature, they all ensure conservation of internal energy within the kernel.
\cite{price08}, \cite{price12a} and \cite{valdarnini12} propose the diffusion of internal energy, while \cite{read12} propose the diffusion of entropy.
\cite{wadsley08} propose a first mixing formulation to resolve the differences in entropy profiles within cosmological comparison simulations \citep{frenk99} between grid and SPH codes.
The diffusion coefficient is approximately proportional to $\alpha^{c}v^\mathrm{sig,c}x_{ij}$ and the numerical coefficient $\alpha^{c}$ is commonly treated as constant through space and time.
We adapt the formulation of a spatially varying coefficient of \cite{tricco13} and additionally calculate a limiter depending on the local hydrodynamical and gravitational states.
We compute the gradient of internal energy as

\begin{equation}(\mathbf{\nabla}u)_{i}=\frac{1}{\rho_{i}}\sum_{j}m_{j}(u_{j}-u_{i})\mathbf{\nabla}_{i}W_{ij}\end{equation}
and approximate the AC coefficient

\begin{equation}\alpha_{i}^\mathrm{c}=\frac{h_{i}}{3}\frac{|\mathbf{\nabla}u|_{i}}{|u_{i}|}\end{equation}
as a measure of noise of internal energy sampling on kernel scale.
The time evolution (i.e. spatially varying SPH discretization of the second-order diffusion equation) of the internal energy for each particle and its neighbours is then given by

\begin{equation}\left.\frac{du_{i}}{dt}\right|_{\mathrm{cond}} =\sum_{j}\frac{m_{j}}{\rho_{ij}}(u_{j}-u_{i})\alpha_{ij}^{c}v^\mathrm{sig,c}_{ij}\overline{F}_{ij},\end{equation}
where we employ the choice of \cite{price08} for signal velocity depending on the pressure gradient of the form

\begin{equation}v^\mathrm{sig,c}_{ij}=\sqrt{\frac{|P_{i}-P_{j}|}{\rho_{ij}}}\end{equation}
and $\alpha^{c}_{ij}=(\alpha^{c}_{i}+\alpha^{c}_{j})/2$ is the symmetrised conduction coefficient, which are individually limited to the interval $[0,1]$.
In the literature several other forms of AC \citep[see eg.][]{wadsley08,valdarnini12} or approaches to the mixing problem \citep[see e.g.][]{hopkins13} have been proposed.

\subsubsection{Gravity limiter}

We note that the amount of AC applied depends on the gradients of internal energy and of pressure.
In the case that the thermal pressure gradient is determined by gravitational forces (i.e. hydrostatic equilibrium) this method would incorrectly lead to unwanted conduction.
In the following, we determine the contribution of hydrostatic equilibrium to the total thermal pressure gradient and present a method to limit the amount of conduction.
Firstly, for every individual active particle, we project the gravitational force $\mathbf{F}^{g}$ onto the hydrodynamical force $\mathbf{F}^{h}_{i}$ and calculate the partial force $\mathbf{F}^{p}_{i}$ of $\mathbf{F}^{h}_{i}$, which is balanced by $\mathbf{F}^{g}_{i}$ to

\begin{equation}\mathbf{F}^{p}_{i}=\frac{\left(\mathbf{F}^{g}_{i}\cdot\mathbf{F}^{h}_{i}\right)}{|\mathbf{F}^{h}_{i}|^{2}}\mathbf{F}^{h}_{i}.\end{equation}
The sign of $\mathbf{F}^{p}_{i}$ depends on the spatial orientation of the force vectors.
Secondly, we subtract/add the partial force $\mathbf{F}^{p}_{i}$ from/to the hydrodynamical force $\mathbf{F}^{h}_{i}$ and obtain $\mathbf{F}^{c}_{i}$, which we call the gravitationally adjusted hydrodynamical force

\begin{equation}\mathbf{F}^{c}_{i}=\mathbf{F}_{i}^{h}+\mathbf{F}_{i}^{p}\end{equation}
and which we use to determine a limitation factor $\delta^{c}_{i}$ for AC

\begin{equation}\delta_{i}^{c}=\left(\frac{\left(\mathbf{F}^{c}_{i}\cdot\mathbf{F}^{h}_{i}\right)}{|\mathbf{F}^{h}_{i}|^{2}}\right)^{q}.\end{equation}
The limiter ensures that AC is only applied to the part of $\mathbf{F}^{h}_{i}$ which is not balanced by $\mathbf{F}^{g}_{i}$.
The exponent $q$ represents a scaling for the aggressivity of the gravity correction.
We limit our correction factor to the interval $[0,1]$ and directly multiply it onto the individual AC coefficients $\alpha_{i}^{c}$.
The limiter performs only as well as the hydrodynamical scheme is able to resolve hydrostatic equilibrium (in the ideal case the angle between force vectors is $180^{\circ}$).
However, in SPH simulations small-scale noise is present at all times within the kernel and thus also in the force vector angles.
The exponent $q$ (applied after the boundary verification) can then be understood to account for the noise in the particle distribution and mimics an opening angle of force vectors.
After extensive studies and performing a variety of test problems, we settle with $q=5$.
The limiter returns zero in the case no hydrodynamical forces are present and one in the case, where no gravitational forces are present.
We are aware that in the presence of strong pressure gradients and rotational forces our approach only marginally limits the amount of AC applied.
However, we did not encounter major problems in our simulations performed with the `new' scheme so far.
Therefore, we assume this issue to be not too important at the present state.

\subsection{Particle wake-up and time-step limiter}\label{sec:wakeup}

GADGET employs individual time-steps for all of the particles to increase computational efficiency.
Thus, the particle population is split into a set of active particles, whose hydrodynamical properties are integrated in the current time-step and a set of inactive particles, which reside on larger time-steps.
These individual time-steps are computed from the local thermodynamical properties of each particle.
However, the splitting between active and inactive computational regions creates problems, where both sets of particles are overlapping.
In the case of a rapid gain in velocity or entropy an active particle can penetrate into a region of inactive particles.
The inactive particles do not notice the sudden presence of the highly dynamical particle and therefore large gradients in the time-steps and unphysical results can occur.
As a treatment we adopt a time-step limiting particle wake-up scheme as proposed and implemented in the GADGET-3 code by \cite{pakmor10} and \cite{pakmor12} with the help of K. Dolag.
It is similar to the time-step limiting scheme described by \cite{durier12} and can be considered an extension of the \cite{saitoh09} mechanism.
Furthermore, our limiter compares signal velocities instead of time-steps and accounts for incorrect extrapolations.
For every active particle, in every time-step, the individual time-steps themselves are re-computed according to

\begin{equation}\Delta{}t_{i}=\frac{Ch_{i}}{v_{i}^\mathrm{sig}},\end{equation}
where $C$ is the Courant factor and $v_{i}^\mathrm{sig}$ the maximum signal velocity (see Section \ref{sec:visc}).
For the calculation of the time-step the maximum of the signal velocity computed between the active particle $i$ and all its neighbour particles $j$ within the entire kernel is used.
GADGET employs a check during the hydrodynamical force computation for large differences in signal velocities (see Eq. (\ref{equ:signal})) within the kernel by evaluating

\begin{equation}v_{ij}^\mathrm{sig}>f_{w}v_{j}^\mathrm{sig},\end{equation}
with a tolerance factor $f_{w}$ corresponding to a wake-up triggering criterion, which captures sudden changes in the pairwise signal velocity.
From our hydrodynamical standard tests we find  $f_{w}=3$ to give reasonable results.
Additionally, the fluid quantities of the recently woken-up particles could have already been predicted half a time-step into the future.
Therefore, the incorrect extrapolation is removed and the contribution from the real time-step added.
These corrections are performed for all particles for which the time-steps are adapted.

%#########################################################################################################
%########################### Hydrodynamic Tests ##########################################################
%#########################################################################################################

\section{Hydrodynamical tests without gravity}

We evaluate the performance and accuracy of the two different SPH implementations with a first set of standard problems.
These first test problems are purely hydrodynamical and do not include gravity or more advanced physics, yet.
Throughout all the test problems we use an adiabatic index of $\gamma=5/3$, the same set of numerical parameters (see Section 2) and we do not specifically tune individual test problems.

\subsection{Sod shock tube}

We consider the Sod shock tube problem \citep{sod78} to study the SPH behaviour in a simple weak shock test.
We set up 630000 particles of equal masses using a relaxed glass file in a three-dimensional periodic box with dimensions $\Delta{}x=140$, $\Delta{}y=1$ and $\Delta{}z=1$.
On the left half side of the computational domain ($x<70$) we initialize 560000 particles with a density of $\rho_L=1.0$ and a pressure of $P_L=1.0$.
On the right half side of the computational domain ($x>70$) we initialize 70000 particles with a density of $\rho_R=0.125$ and a pressure of $P_R=0.1$.

\begin{figure}
\begin{center}
  \includegraphics[width=0.475\textwidth]{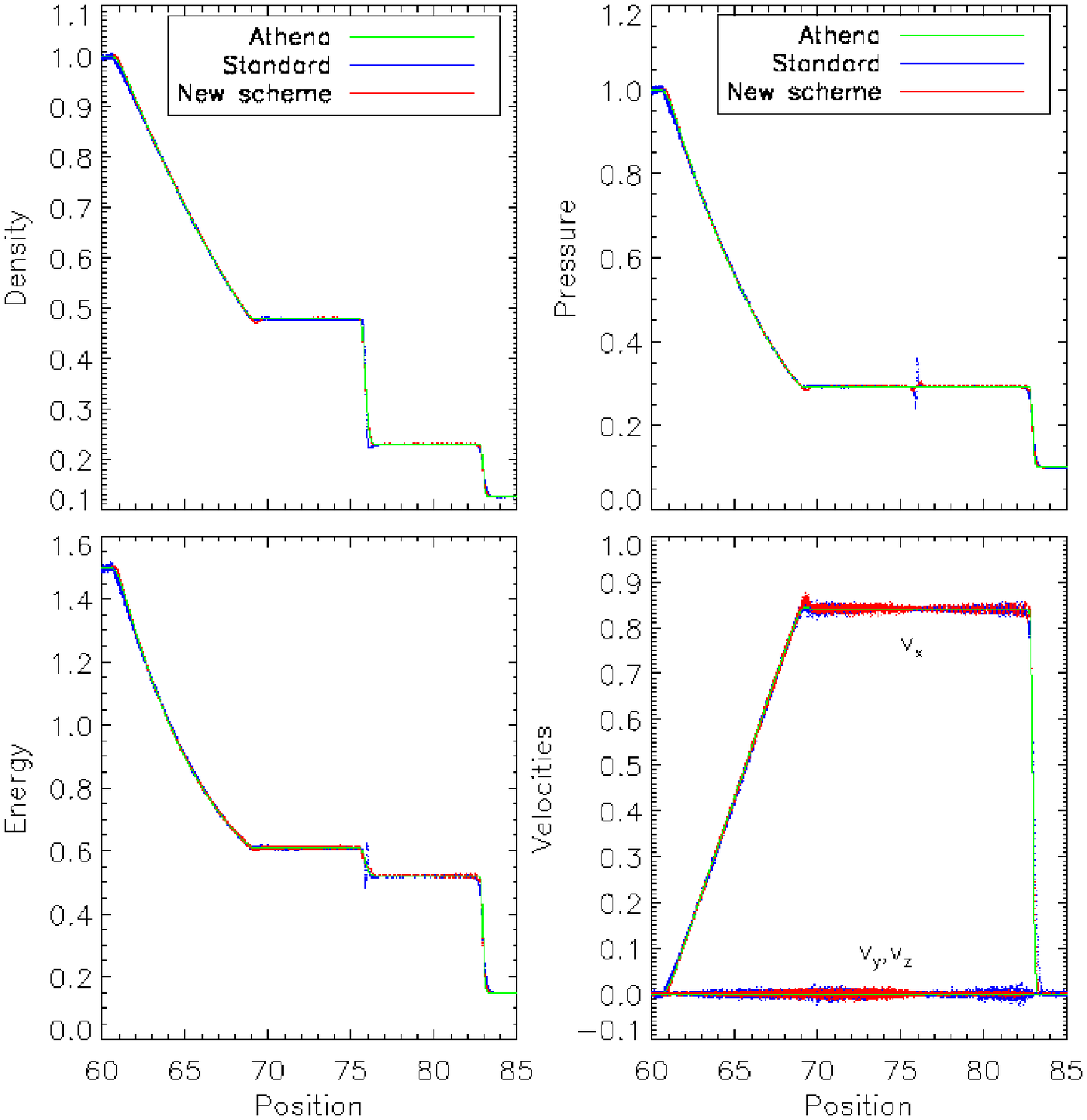}
  \caption{Sod shock tube.
We show the spatial distribution of particles (every $10_{th}$ particle is plotted) for density, thermal pressure, total energy and velocities at time $t=$ 5.0.
Both SPH schemes capture the shock well but with differences as follows.
The `new' scheme converges better in the density estimate and the presence of AC nearly removes the pressure blip at the contact discontinuity.}
  \label{fig:test_sod}
\end{center}
\begin{center}
  \includegraphics[width=0.475\textwidth]{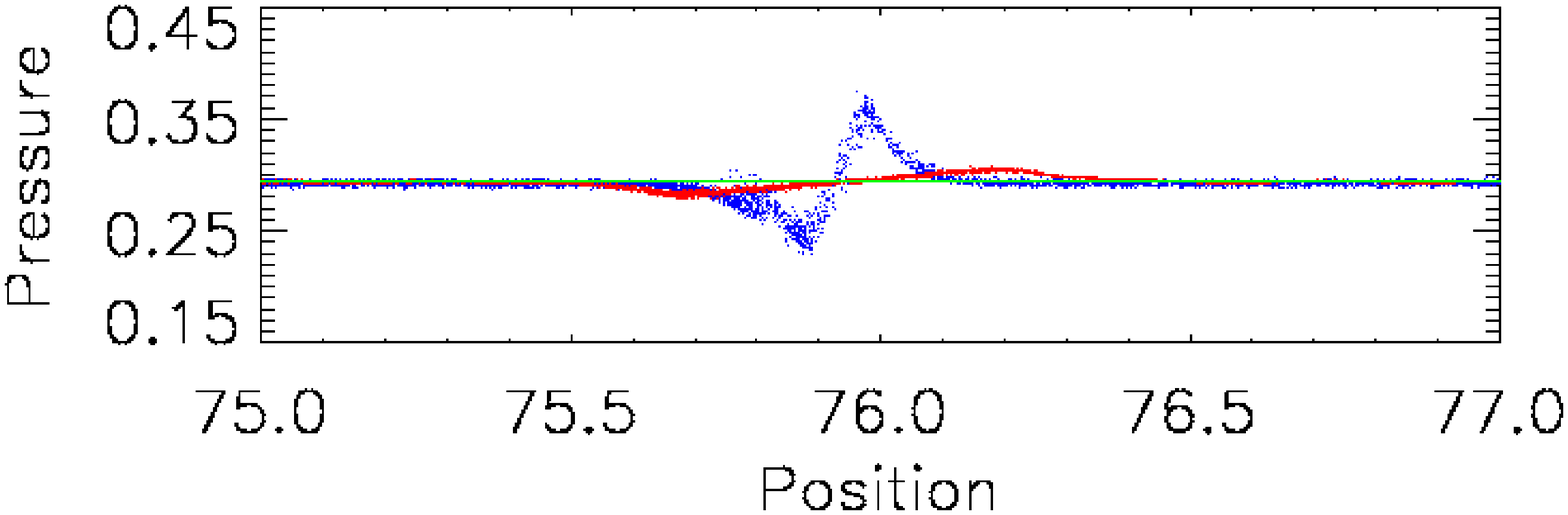}
  \caption{Sod shock tube.
We show a zoom-in on the pressure blip (see Fig. \ref{fig:test_sod}, upper right panel) at time $t=$ 5.0.
Only a small residual of surface tension is left.}
  \label{fig:sod_inlay}
\end{center}
\begin{center}
  \includegraphics[width=0.475\textwidth]{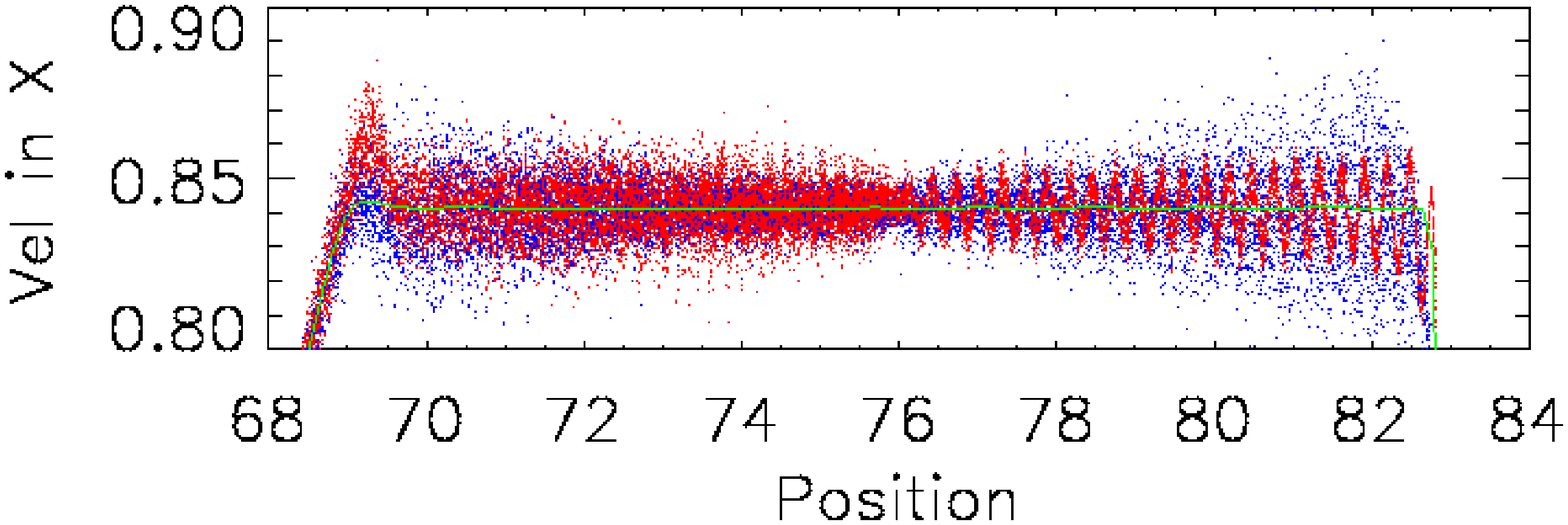}
  \caption{Sod shock tube.
We show a zoom-in on velocity in $x$-direction (see Fig. \ref{fig:test_sod}, bottom right panel) at time $t=$ 5.0.
Besides some post-shock ringing, the post-shock noise is smaller with the 'new' scheme.}
  \label{fig:sod_inlay2}
\end{center}
\end{figure}

Fig. \ref{fig:test_sod} shows the results of the test problem at time $t=5.0$.
In general, both SPH schemes agree fairly well with a reference solution (green line) obtained with the ATHENA code \citep{stone08} but we note the following differences.
In the `standard' scheme (blue dots), the discontinuity in internal energy results in a 'blip' of pressure (see Fig. \ref{fig:sod_inlay}) and energy, which corresponds to the artificial spurious surface tension of SPH.
The issue of the 'blip' has been discussed for a long time \citep[see e.g.][]{monaghan92}.
In the `new' scheme (red dots), AC promotes mixing, resolves the discontinuity, regularizes the pressure and provides a treatment of the 'blip'.
A closer look at individual particles (see Fig. \ref{fig:sod_inlay2}) shows that the pure noise in velocity of particles behind the shock front is lower, which is a direct result of the improved prescription of AV.
However, reducing the viscosity gives rise to post-shock ringing.
Additionally, the change of kernel function improves the sampling quality of the fluid and yields a smoother estimation of density.
At last, the time-step limiter is of little importance due to the weak shock in this test.
However, as seen in Fig. \ref{fig:sod_inlay} the 'blip' is not completely removed and this is where some residual surface tension shows up.

\subsection{Sedov blast}

We consider the Sedov blast problem \citep{sedov59} to study the SPH behaviour in a simple ultrasonic strong shock test.
We set up 130$^3$ particles of equal masses using a relaxed glass file in a three-dimensional periodic box with dimensions $\Delta{}x=\Delta{}y=\Delta{}z=6$ kpc.
In the entire computational domain we initialize the particles with a density of $1.24\times10^{6}$ M$_\odot$ kpc$^{-3}$ and one Kelvin as temperature.
At the centre of the box we point-like distribute the energy $E=6.78\times10^{53}$ erg to mimic a supernova explosion among the nearest 102 particles.

\begin{figure}
\begin{center}
  \includegraphics[width=0.475\textwidth]{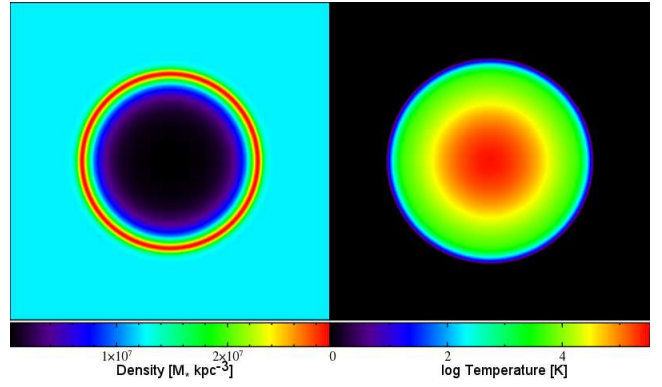}
  \caption{Sedov blast wave.
We show thin slices through the centre of the computational volume at time $t=$ 0.03 of the test performed with the `new' scheme.
The shock front is clearly visible in the gas density (left panel) as well as the temperature (right panel).}
  \label{fig:sedov_contours}
\end{center}
\end{figure}

\begin{figure}
\begin{center}
  \includegraphics[width=0.475\textwidth]{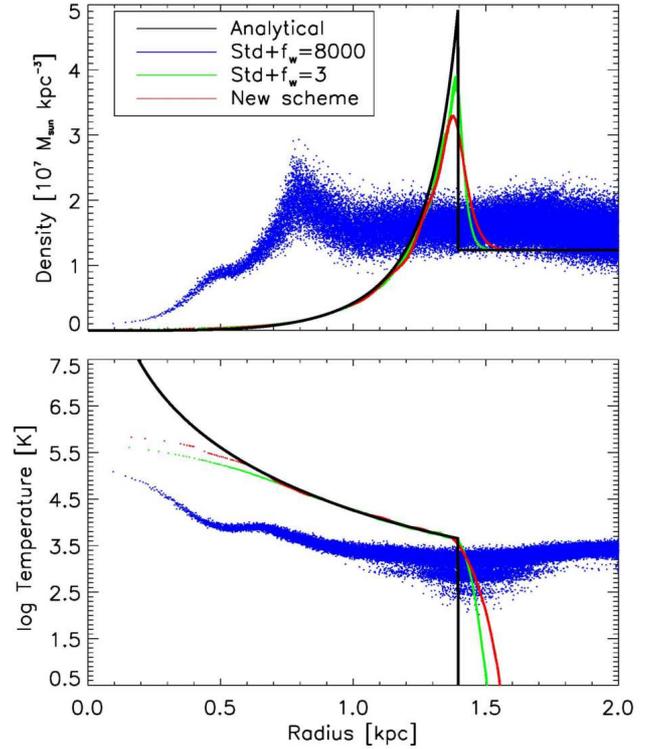}
  \caption{Sedov blast wave.
We show the radial distribution of particles (every $5_{th}$ particle is plotted) at time $t=$ 0.03 with a time-step limiting criterion of $f_w=3$.
We have performed the `standard' run with a time-step limiting criterion of $f_w=8000$ (green lines, otherwise no meaningful comparison can be performed) and also $f_w=3$.
The classic `standard' scheme (green lines) fails to capture the shock, while the `new' scheme (red lines) captures the position and evolution of the shock front much better compared to the analytical solution (black lines).
This test shows the importance of the time-step limiter.}
  \label{fig:test_sedov}
\end{center}
\end{figure}

Fig. \ref{fig:sedov_contours} shows thin slices through the centre of the simulation box and Fig. \ref{fig:test_sedov} the corresponding particle distribution at time $t=0.03$.
Furthermore, we perform the `standard' scheme test also with the time-step limiter ($f_w=8000$) because of the very strong shock (Mach $\gg$ 100) of the blast and otherwise any comparison will fail.
Without the limiter, shocked particles penetrate into quiescent regions causing a highly distorted fluid sampling, which results in an incorrect solution leading to an incorrect propagation of the shock front \citep[see discussions in][]{saitoh09,durier12} and corresponding smoothing.
The entire 'new scheme' reproduces the analytical solution (black line) very well, with the `standard' run (green dots) totally failing, and the `new' run (red dots) capturing the position, density and temperature of the shock fairly well.
In addition, we show a partially improved `standard' run (blue dots), where we enabled the time-step limiter but nothing else.
This run also yields reasonable good results in this test, but as we see later comes short in other tests.
We see that the `new' scheme yields a smooth distribution of particles within the central region and therefore a well-resolved, but smoothed due to AC, temperature solution.

\subsection{Keplerian ring}

We consider the Keplerian ring problem \citep{cartwright09,cullen10} to study the SPH behaviour in a simple rotating and shearing test problem.
We set up 20000 particles of equal masses sampling a two-dimensional ring with a Gaussian surface density profile with a peak at radius $R=15.0$ kpc and a standard deviation of $\sigma=2.0$ kpc.
For numerical reasons we initialize the distribution in concentric shifted circles and not in a random fashion \citep{cartwright09}.
We set the particles on Keplerian orbits around a central $10^9$ M$_\odot$ point mass with a rotation period of $t=2\pi$.
We choose the sound speed orders of magnitudes smaller than the orbital velocity to ensure thermal stability of the ring.
In contrast to our common set of numerical parameters, we start without a minimal value of AV because it would immediately trigger instability.
In the absence of AV the ring should be stable.

\begin{figure}
\begin{center}
  \includegraphics[width=0.475\textwidth]{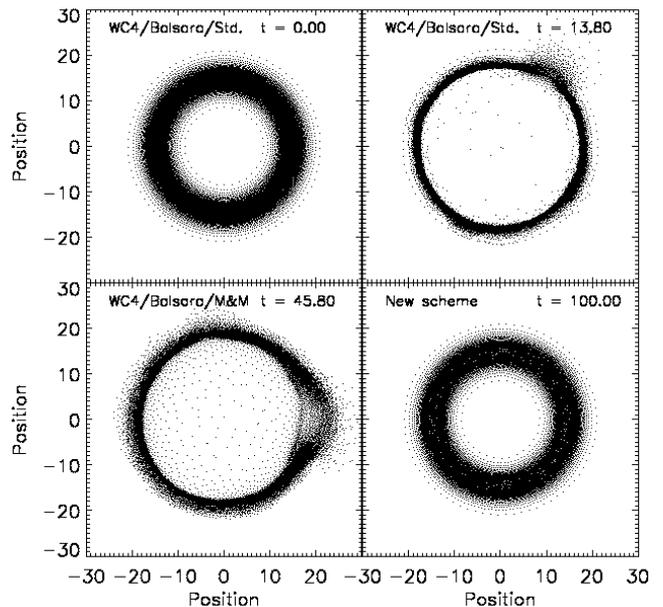}
  \caption{Keplerian Ring.
We show the initial set-up (top left panel) as well as the results of three different AV methods.
For a fair comparison we only vary the AV scheme and none of the other SPH modifications.
The importance of AV becomes clear as high amounts of viscosity lead to numerical accretion of particles onto the central point mass.
Because angular momentum is conserved, the ring breaks up and an instability develops.
With a low-order Balsara limiter, neither the standard SPH viscosity (top right panel) nor the M\&M viscosity (bottom left panel) are able to preserve the ring.
The `new' scheme (bottom right panel), which has a the time-dependent AV coupled with a high-order limiter, is able to preserve the ring to even very late times.}
  \label{fig:test_ring}
\end{center}
\end{figure}

Fig. \ref{fig:test_ring} shows the results of the test problem at the times of onset of runaway instability.
We perform all test runs with the WC4 kernel in order to exclude possible effects caused by the smoothing scale of kernel sizes and differential estimators.
Due to the highly sub-sonic nature and the absence of strong shocks, the impact of AC and the time-step limiter is negligible.
The initially stable ring (top left panel) evolves as follows for different implementations of AV.
In the `standard' scheme (top right panel), the ring is only stable for about two dynamical times, before the instability has fully developed and the ring breaks up.
Also, the Balsara limiter does not succeed in limiting AV because of the insufficient calculation of vorticity.
In the 'M\&M' scheme (bottom left panel) we use the implementation of a low-viscosity scheme initially proposed by \cite{morris97} and implemented into GADGET by \cite{dolag05}.
Their scheme uses a time-dependent evolution of numerical AV coefficient $\alpha^{v}_{i}$ to suppress AV in the absence of shocks and manages to keep the ring stable for about seven dynamical times.
However, the M\&M scheme requires a minimum value of AV and also uses a low-order estimator for vorticity, which leads to the ring break-up.
At last, we show the results of the `new' scheme (bottom right panel), which we used without a minimum value for AV.
This scheme uses a high-order estimator of velocity gradient matrix, which results in a very accurate calculation of divergence and vorticity.
Therefore, also the computation of the Balsara shear flow limiter is very accurate and suppresses AV completely within the entire ring structure.
We do not note an artificially induced transport of angular momentum and orbital changes of test particles.
Consequently, the ring remains stable for many dynamical times and the initial Gaussian surface density distribution is preserved until we stopped the simulation.

\subsection{Cold blob test}

We consider the blob test \citep{agertz07,read10} set up with publicly available initial conditions\footnote[1]{http://www.astrosim.net/code/doku.php} to study the SPH behaviour in a test problem with interacting gas phases and surfaces.
We initialize 9641651 particles of equal masses using a relaxed glass file in a three-dimensional periodic box with dimensions $\Delta{}x=10$, $\Delta{}y=10$ and $\Delta{}z=30$ in units of the cold cloud radius.
A cold cloud is centred at $x,y,z=5$ and travels at a Mach number of $M=2.7$.
The background medium is set-up ten times less dense and ten times hotter than the cloud.
Spherical harmonics are used to seed large-scale perturbations onto the surface of the cloud.
Because of the low Mach number shocks we expect the time-step limiter to be only of minor importance.

\begin{figure*}
\begin{center}
  \includegraphics[width=0.95\textwidth]{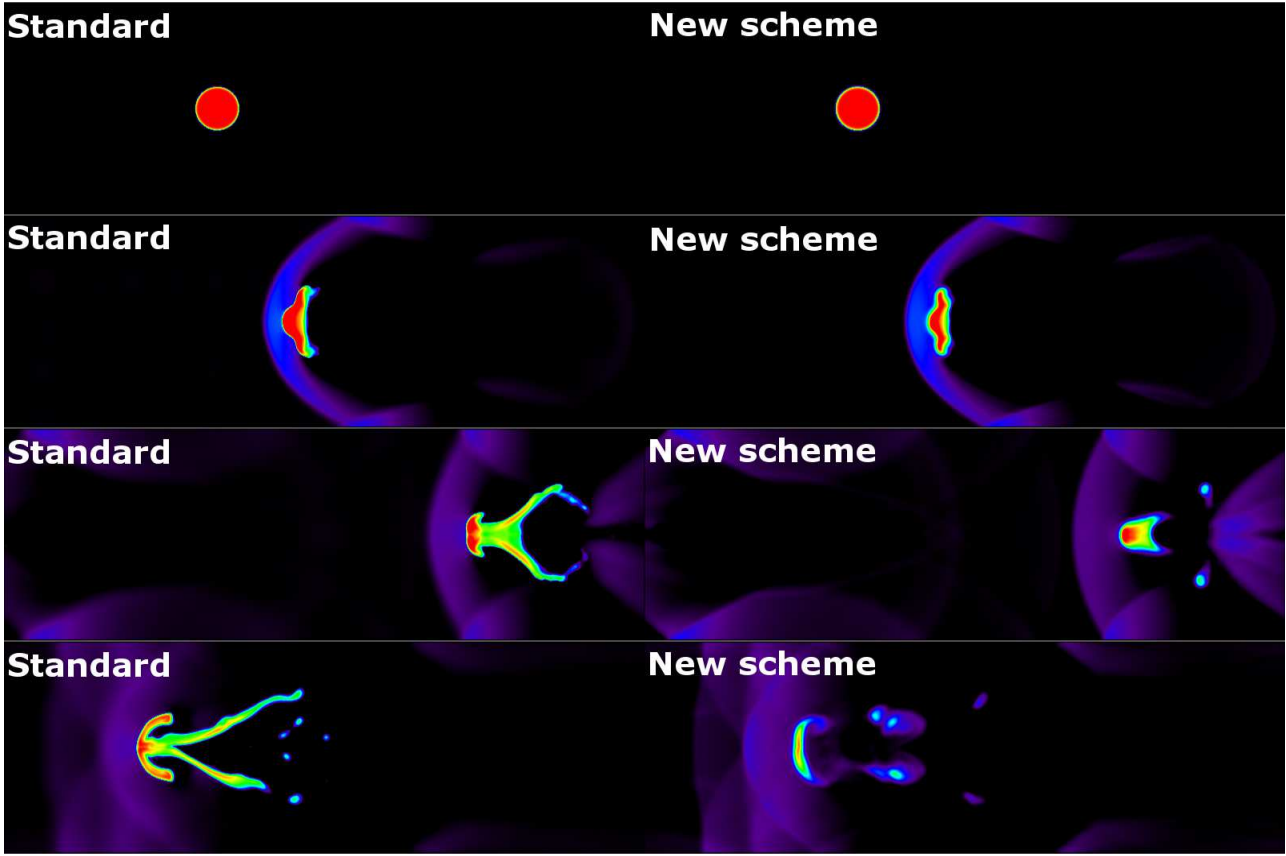}
  \caption{Blob test.
We show thin slices of gas density through the centre of the computational domain at times $t=$ 0.0, 3.0, 6.0 and 10.0.
In the `standard' scheme, numerical surface tension prevents mixing between cold and hot phases leading to an artificial stretching of the cloud and an unphysical solution.
In the `new' scheme, AC helps to promote cloud dissociation; however, some residual surface tension remains left.
Furthermore, the shock structures throughout the box are more defined and better resolved.}
  \label{fig:test_blob}
\end{center}
\end{figure*}

Fig. \ref{fig:test_blob} shows thin slices through the density structure at various times.
In the `standard' scheme, the cold gas cloud is prevented from dissociating by the follow major effect \citep[see also e.g.][]{agertz07}.
The presence of artificial surface tension confines the blob of cold gas.
This is clearly visible by the numerically induced stretching of the cloud.
Cold material, which should have been mixed into the ambient hot medium is confined within an elongated structure.
In the `new' scheme, the presence of AC promotes the mixing with external ambient medium.
We also note that the `new' scheme resolves different shock structures propagating through the box both more accurately and smoothly.

\begin{figure}
\begin{center}
  \includegraphics[width=0.475\textwidth]{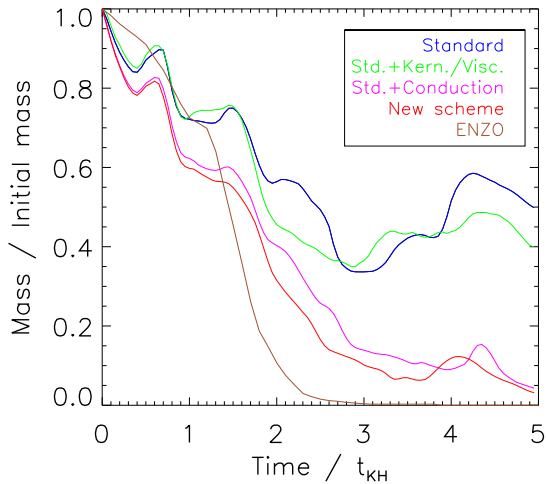}
  \caption{Blob test.
We show the fraction of cold gas as a function of time for four different SPH schemes.
In the `standard' scheme, numerical surface tension prevents mixing between the cold and hot phases.
The importance of AC becomes clear as it promotes mixing between gas phases, which allows a dissociation of the cloud.
Only some residual surface tension is left.
Comparing to a run performed with the ENZO grid code, we find our AC scheme to model mixing in a conservative way.}
  \label{fig:test_blob_loss}
\end{center}
\end{figure}

Fig. \ref{fig:test_blob_loss} illustrates the dissipation of the gas cloud by tracking the time evolution of cold blob mass.
We associate \citep[see also][]{agertz07} particles with the cold cloud with a temperature criterion of $T<0.9\cdot{}T_\mathrm{ext}$ (in contrast to the external ambient medium) and a density criterion of $\rho>0.64\cdot{}\rho_\mathrm{cl}$ (in contrast to the initial density of the cloud).
In the `standard' scheme (blue line), only half of the cold gas mass is mixed into the hot ambient medium over five dynamical times.
The effects of the improved AV and WC4 kernel (green line) are negligible.
The major impact and contribution to cloud dissociation is made by AC (pink line) and the corresponding introduced mixing process.
The results are close to a test run performed with the ENZO \citep{bryan14} code in a comparable set-up, which we took from \cite{hopkins13}.
However, some residual surface tension remains.

\subsection{Kelvin-Helmholtz instability}

We consider the Kelvin-Helmholtz instability \citep{agertz07,read10} from publicly available initial conditions\footnotemark[1] to study the SPH behaviour in a simple shearing instability test.
We set up 1548288 particles of equal masses using a cubic lattice in a three-dimensional periodic box with dimensions $\Delta{}x=256=\Delta{}y=256$ and $\Delta{}z=16$ kpc, which is centred around ($0,0,0$).
In the central half of the box ($|y|<64$) we initialize 512000 particles with a density of $\rho_1=6.26\cdot{}10^{3}$ M$_\odot$kpc$^{-3}$, temperature of $T_1=2.5\cdot{}10^{6}$ K and a velocity in $x$-direction of $v_1=-40$ km/s.
In the outer half of the box ($|y|>64$) we initialize 1036288 particles with a density of $\rho_2=3.13\cdot{}10^{2}$ M$_\odot$kpc$^{-3}$, temperature of $T_2=5.0\cdot{}10^{6}$ K and a velocity in $x$-direction of $v_2=40$ km/s.
To trigger the instability, we perturb the velocity in $y$-direction with a mode of wavelength 128 kpc and amplitude of 4 km/s at the boundary layer that is exponentially damped towards the upper and lower edge of the box.

\begin{figure*}
\begin{center}
  \includegraphics[width=0.95\textwidth]{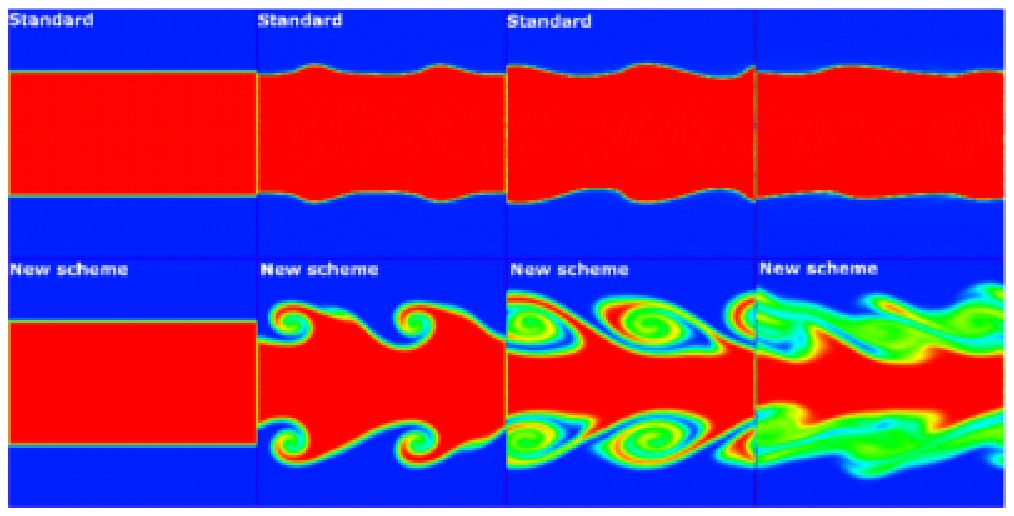}
  \caption{Kelvin-Helmholtz instability.
We show thin projections of specific entropy through the centre of the computational domain at times $t=$ 3.0 and 6.0 (dynamical timescale $t_\mathrm{KH}\approx3.4$).
In the `standard' scheme, numerical surface tension as well as too much AV prevent the instability to develop and lead to an unphysical laminar behaviour of the fluid.
In the `new' scheme, our formulations of AV and AC promote the formation of roll-ups and onset of instability, while at late stages diffusion is dominating.}
  \label{fig:test_kelvin}
\end{center}
\end{figure*}

Fig. \ref{fig:test_kelvin} shows thin projections through the specific entropy structure of the test problem at various times.
In the `standard' scheme, the fluid evolves in a laminar fashion and the growth of perturbations is totally suppressed by the artificial surface tension confining the central gas stream and large amounts of AV damping velocity perturbations \citep[see also e.g.][]{agertz07,price08}.
In the `new' scheme, the high-order Balsara shear limiter successfully limits AV and allows large-scale perturbations to develop two prominent roll-ups.
Additionally, AC nearly removes the artificial surface tension between the two gas phases and promotes mixing within the roll-ups.
The entire test set-up does not evolve completely symmetric because of small secondary perturbations caused by the initial set-up on a cubic lattice.
Most importantly, the high-order AV and AC prove crucial for this test problem, while the WC4 kernel and time-step limiter are of less importance.
At the late stages, in this set-up the `new' scheme is dominated by diffusion.

\subsection{Decaying Subsonic Turbulence}

Recent comparisons of standard SPH implementations with static and moving mesh grid codes have sparked a debate about the capabilities of SPH to model subsonic turbulences \citep[see e.g.][]{bauer12,price12b,hopkins13,hopkins15a}.
We study the behaviour of our `new' scheme in Idealized simulations of decaying subsonic turbulence.
In particular, we are interested in the effective viscosity of the two schemes and the behaviour of the 'SPH noise' under conditions appropriate to galaxy formation and cluster simulations, i.e. non-isothermal, decaying motions from solenoidal and compressive modes. As most baryons on cosmological scales are in weakly collisional plasmas, numerical models should aim to minimize viscosity where possible \citep[see e.g.][]{brunetti07}.

\subsubsection{Grid and particle conversion procedures}

We set up $512^3$ particles of equal masses within a periodic box of side length 3 Mpc/h using carefully relaxed SPH glass files to minimize spurious initial kinetic energy.
Subsequently, we define a velocity field on a grid of the same resolution in $k$-space by sampling a spectral distribution using the Box-Mueller method.
The velocity field is transformed back to real-space using a Fast Fourier Transformation (FFT), normalized such that the average velocity is of the desired Mach number.
The velocities from the grid are transferred to the particle distribution using the nearest grid point (NGP) sampling kernel.

To assess the impact of random motions near the resolution scale, we need to measure the velocity power spectrum  within the SPH kernel. However, the accurate estimation of the velocity power spectrum of a particle distribution close to the Nyqvist frequency is non-trivial, because of aliasing of the velocity power by the binning kernel \citep[see discussions in][]{jing05,jasche09,cui08}. This can be compared to a problem in signal processing, where SPH represents an analogue signal and a grid a digital signal representation of it. Aliasing is strongest at the smallest scales/largest modes, where the velocity power on the particles is modified by the shape of the binning kernel in configuration-space.
To understand this effect and compare binning kernels, we take initial conditions with a full Kolmogorov power spectrum ($P_\mathrm{k} \propto k^{-11/3}$) without performing a simulation and directly bin the velocity back to a grid using different kernels.
After a forward fast Fourier transformation we radially average the velocity power in $k$-space in 32 logarithmic bins.

\begin{figure}
\begin{center}
  \includegraphics[width=0.475\textwidth]{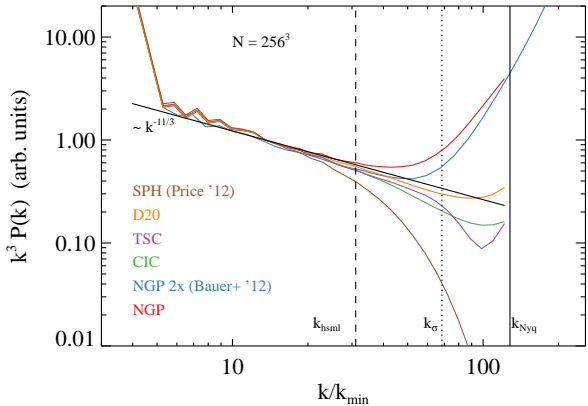}
  \caption{Comparison of different particle to grid binning methods acting on the same particle distribution sampling a Kolmogorov velocity power spectrum in three dimensions (black line).
We show the NGP kernel (red line), the twice oversampled NGP (blue line), the CIC (green line), the TSC (violet line), the D20 (orange line) and the SPH WC4 (brown line).
The vertical lines indicate the wave numbers corresponding to the Nyqvist frequency (solid line), the WC4 compact support (dashed line) and the WC4 smoothing scale (dotted line).}
  \label{fig:turb_binning}
\end{center}
\end{figure}

Fig. \ref{fig:turb_binning} shows the resulting power-spectra where the black line represents the original power spectrum.
We show the kernels:  Nearest Grid Point (NGP, red line), Cloud in Cell (CIC, green line), Triangular Shaped Cloud (TSC, violet line), Daubechies scaling function of 20th order (D20, orange line), and the WC4 SPH kernel with 200 neighbors (SPH, brown line) \citep{hockney88,daubechies92,dehnen12}.
We also show the NGP with two times oversampling (blue line), which was used in \cite{bauer12}.
As vertical lines we show the Nyqvist frequency $k_\mathrm{Nyq} = N k_\mathrm{min}$, the WC4 smoothing scale $k_\sigma$ and the WC4 kernel compact support $k_\mathrm{hsml} = \pi/{h_\mathrm{sml}}$ \citep{dehnen12}.

During the binning process, the SPH kernel function conserves density to machine precision but not energy, i.e. binning with the SPH kernel is a diffusive process.
The other kernels behave opposite, they conserve mass, scalar velocity and energy to less than one per cent but not SPH density and volume.
Fig. \ref{fig:turb_binning} clearly shows that the D20 wavelet kernel minimizes aliasing for sufficiently homogeneous particle distributions \citep{cui08}.
Our comparison also resolves the differences found in \cite{bauer12} and \cite{price12b}, who use the twice oversampled NGP kernel and the standard SPH kernel, respectively.
Prior studies based on the NGP kernel binning over-estimated the SPH noise, while SPH kernel based binning suppressed the real noise by aliasing.
We conclude that all kernels except the D20 show substantial aliasing and it seems hard to draw definitive conclusions from simulation results under this condition. Thus we make it our fiducial choice for this study.
We note that in the presence of strong gradients in density the SPH kernel remains the only viable choice to obtain binned quantities, because it is the only kernel in our comparison that guarantees a non-negative non-zero density in the entire simulation at all grid resolutions.
This works reasonably well for a physical interpretation of velocity power-spectra, because motions below the SPH smoothing scale are caused by numerical effects \citep{price12b}.

\subsubsection{Spectral evolution of turbulence}

\begin{figure*}
\begin{center}
  \includegraphics[width=0.85\textwidth]{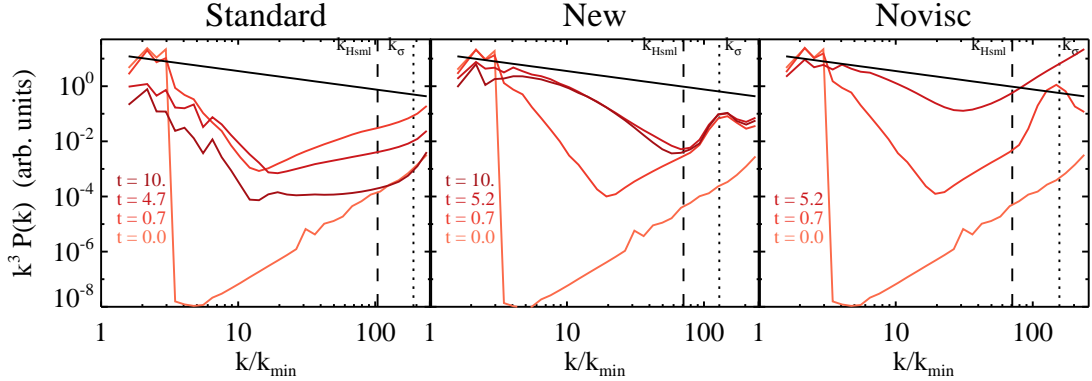}
  \caption{Decaying subsonic turbulence.
We show the build-up and decay of velocity power spectra for different schemes.
The colours illustrate the time evolution of the spectra (sound-crossing time of about $t_s=7.0$.
We initially distribute energy on the largest modes, which then develops a spectral distribution.
In the `standard' scheme (left panel) turbulent motions are almost completely suppressed by destructive impact of AV.
In the `new' scheme (middle panel) turbulent motions develop and a turbulent cascade is built.
The diffusive character of AV is significantly changed and the velocity field as well as the kinetic energy are preserved.
The spectra are then compared to a simulation without viscosity (right panel).}
  	\label{fig:turb_spectra}
\end{center}
\end{figure*}

\begin{figure*}
\begin{center}
  \includegraphics[width=0.95\textwidth]{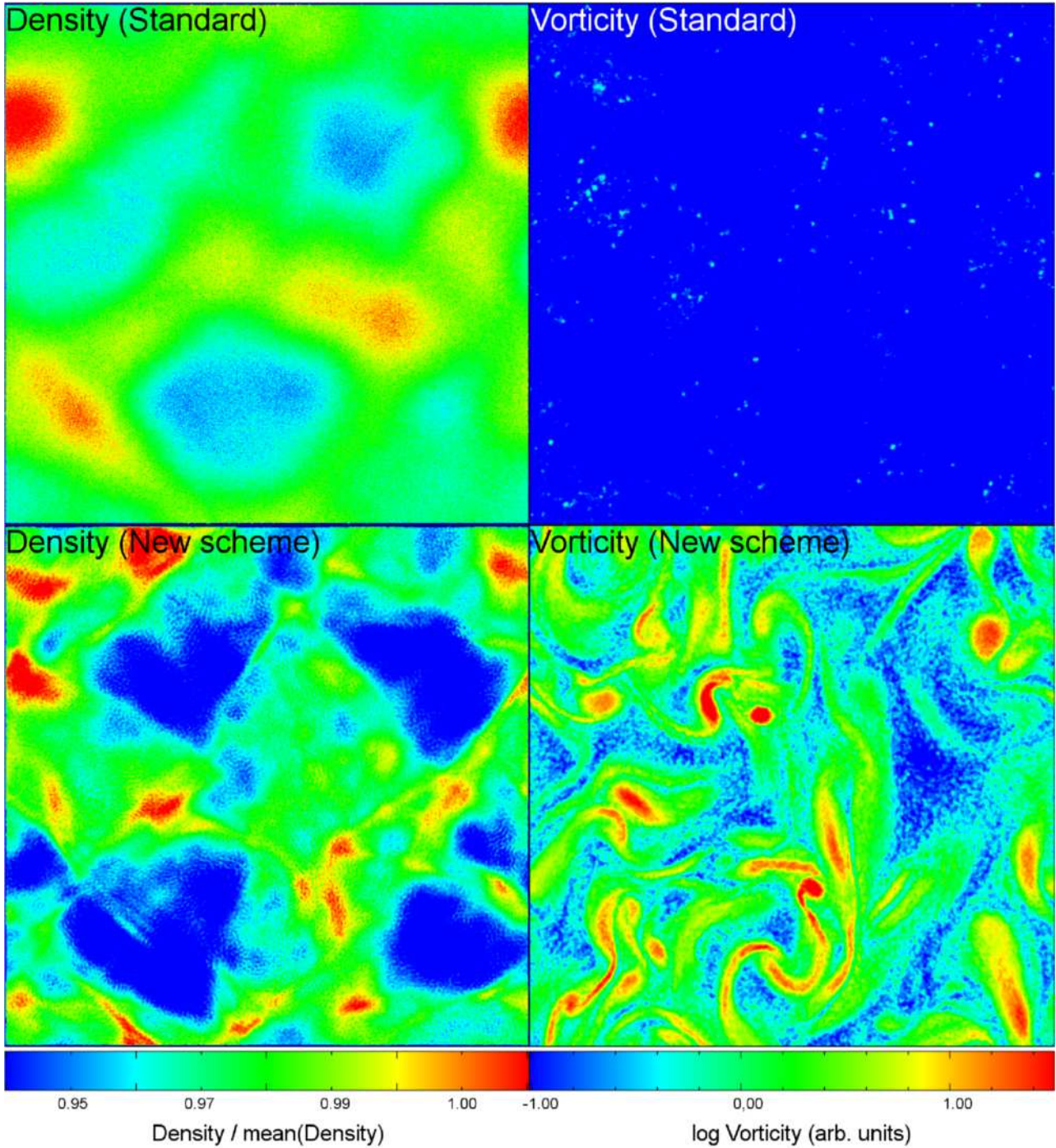}
  \caption{Decaying subsonic turbulence.
We show thin slices through the centre of the simulation box after one sound-crossing time for both schemes.
The panels correspond to gas density (left panel) and vorticity (right panel).
The velocity field shows well developed turbulence consisting of compressive and shearing motions.
The `new' scheme is able to more accurately compute vorticity and suppress AV with the high-order Balsara limiter.}
  \label{fig:turb_slices}
\end{center}
\end{figure*}

\begin{figure}
\begin{center}
  \includegraphics[width=0.475\textwidth]{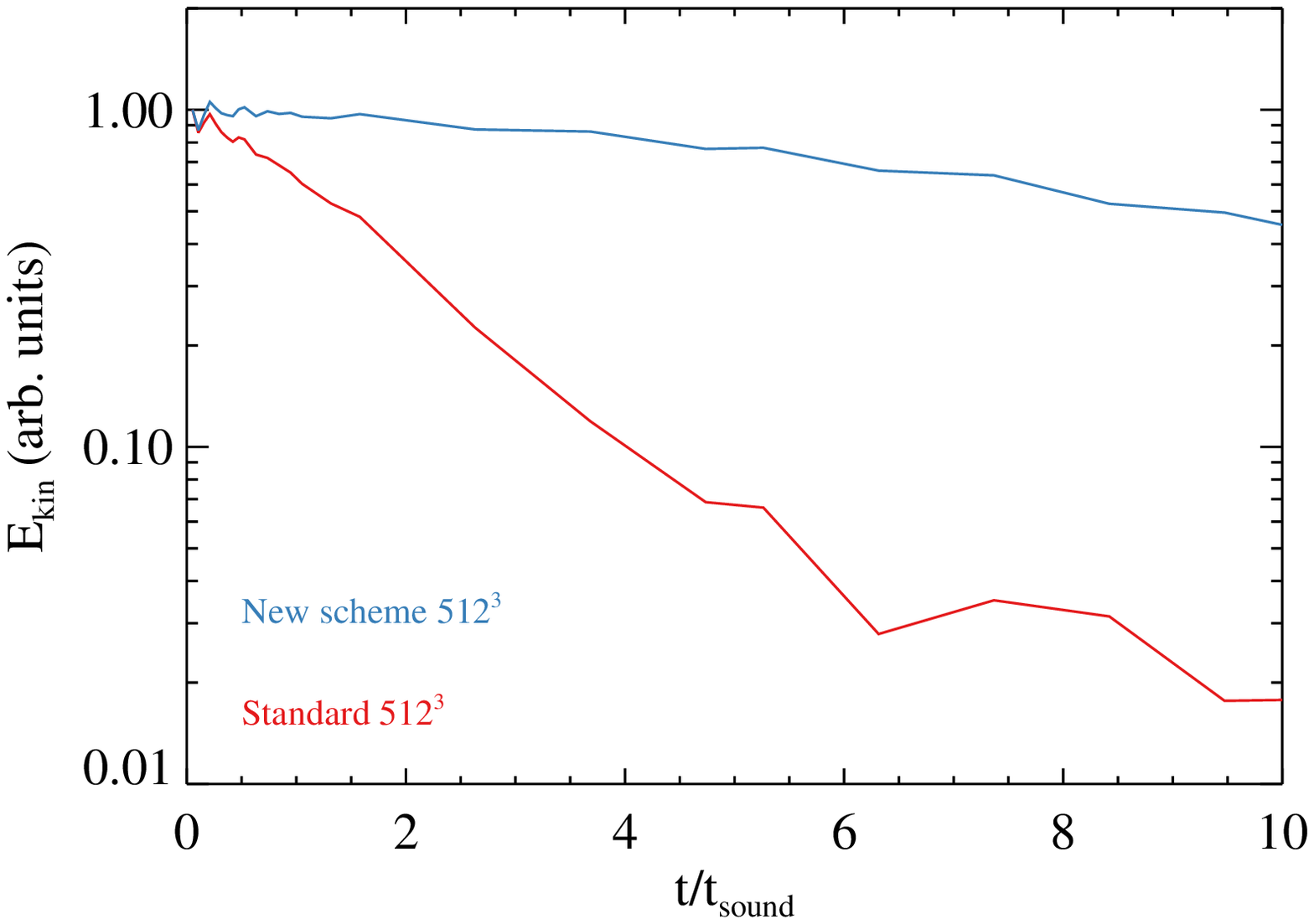}
  \caption{Decaying subsonic turbulence.
We show the total kinetic energy in the simulation box over time.}
  \label{fig:turb_energy}
\end{center}
\end{figure}

To compare the `standard' and the `new' schemes we consider decaying turbulence within a periodic box.
We seed compressive and solenoidal modes in the range of $k \in [1.6, 3.1]$, to obtain initial conditions appropriate for the galaxy and cluster environment, where turbulence is injected by merger infall on the scale of the halo core radius.
We normalize the velocity fluctuations in the box such that the average velocity equals a Mach number of $M = 0.1$ and we do not time-average spectra.

Fig. \ref{fig:turb_spectra} (left and middle panels) shows the time evolution of velocity power-spectra for the `standard' scheme (left panel) and the `new' scheme (middle panel).
Here we also show the scale of the SPH kernel compact support (black vertical line) and the kernel smoothing scale (dotted vertical line).
In-line with previous studies \citep[][their fig. 12]{bauer12}, the `standard' scheme does not develop a turbulent cascade and damps kinetic energy very quickly.
Our `new' scheme develops a cascade  at large scales (small $k$) but then shows a depression of kinetic energy close to the kernel scale.
This, again, is in-line with prior studies \citep{hopkins13}.
The damping of the spectrum at the later times appears self-similar, i.e. the shape of the spectrum does not change as energy decreases.
Inside the kernel the typical build-up of thermal motions around the smoothing scale $k_\sigma$ can be observed, but scales outside the kernel are not affected.

In order to understand if the cause of the velocity depression at $k \approx 10$ is caused by the formulation of AV we perform a test-run without any viscosity (Fig. \ref{fig:turb_spectra}, right panel).
Throughout the whole evolution, the spectrum at the smallest $k$ follows the Kolmogorov scaling, as expected.
Once the turbulent cascade is established, the spectrum turns over at increasingly smaller scales, which is equivalent to an isotropization of kinetic energy inside the kernel and subsequent filling of larger scales with isotropic motions.
This follows from the fact that SPH particles are subject to the pair-wise repulsive force \citep{price12a}, and hence behave like a thermal gas below the kernel scale.
Eventually, such a system will show a flat power spectrum as expected from the second law of thermodynamics.
In our simulation, we also observe the depression outside the smoothing scale, which seems to be an intrinsic feature of SPH related to an energy transfer from outside the kernel to smaller scales, and not related to our formulation of AV.

In our `new' scheme including AV, thermal motions are well controlled inside the kernel scale, commonly referred to as 'kernel noise'.
We argue that these motions are not spurious, because no additional energy is retained in them, as SPH is fully conservative and the spectrum decays roughly in a self-similar manner.
If we define the kernel smoothing scale ($k_\sigma$) as the dissipation scale intrinsic to SPH, the `new' scheme does not show a bottle-neck effect as found in grid codes at adjacent larger scales but a depression, roughly in the same range in $k$.
This is in line with the results shown by \cite{hopkins15a}, whose code uses a Riemann solver to formulate noise-free AV on scales of the inter-particle separation to obtain grid-code behaviour.
We note that the difference in dissipation scale ($d_\mathrm{min}$ versus $\sigma_\mathrm{kernel}$) translates into more resolution elements required by SPH compared to Eulerian methods, i.e. slower convergence. 

Fig. \ref{fig:turb_slices} shows thin slices through the centre of the simulation box after one sound-crossing time.
We visualize gas density (left panel) and vorticity (right panel).
It can be clearly seen that our `new' scheme resolves compressive and shearing velocity motions better than the `standard' scheme
The high-order derivatives of velocity lead to a more accurate estimation of vorticity and thus, limit the impact of AV and preserve kinetic energy and turbulent motions.
Furthermore, the difference in AV and velocity dissipation between the two schemes becomes strikingly evident in the time evolution of kinetic energy (Fig. \ref{fig:turb_energy}).
The `standard' scheme dissipates $90 \%$ of the kinetic energy budget within four sound-crossing times, while the `new' scheme preserves energy better by a factor of 5.

We conclude that our code performs comparably to modern implementations of SPH \citep{hopkins13,price12a}, even in the case of non-isothermal compressive and solenoidal decaying turbulence found in cosmological simulations.
We show that the disagreement between \cite{bauer12} and \cite{price12a} is largely caused by technical differences, to solve it we propose a solution based on the D20 binning kernel.
We also show that the downturn in the velocity power spectrum is not caused by the AV implementation.

%#########################################################################################################
%########################### Hydrodynamic Tests ##########################################################
%#########################################################################################################

\section{Hydrodynamical tests with gravity}

We continue to evaluate the performance and accuracy of the two different SPH implementations with a second set of standard problems.
These second tests include hydrodynamical as well as gravitational forces and also take a cosmological time integration into account.

\subsection{Hydrostatic sphere}

We consider a sphere in hydrostatic equilibrium to study the SPH behaviour in combination with gravity in an ideally stable system.
We set up 88088 dark matter particles with individual masses of $2\cdot{}10^{9}$ M$_\odot$ and 95156 gas particles with individual masses of $4.75\cdot{}10^{8}$ M$_\odot$.
The total mass of the sphere is $2.2\cdot{}10^{14}$ M$_\odot$ and we use vacuum boundary conditions and a gravitational softening length of 12 kpc.
We set up the initial equilibrium conditions following \cite{komatsu01}, as described in \cite{viola08}.
We evolve the sphere adiabatically and do not include cooling or heating mechanisms in this test.

\begin{figure*}
\begin{center}
  \includegraphics[width=0.95\textwidth]{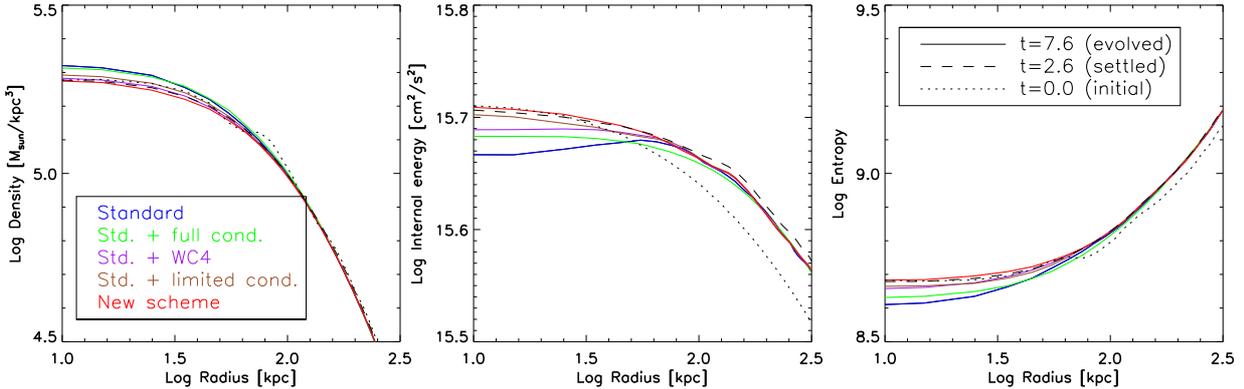}
  \caption{Sphere in hydrostatic equilibrium.
We show radial profiles of density (left panel), internal energy (middle panel) and entropy (right panel) at three different times.
At first, the initial conditions (dotted lines) must settle into hydrostatic equilibrium (dashed lines), which then remains stable for an extended period of time (solid lines).
The stability of the sphere is determined by the differences occurring between the settled (only shown for the `new' scheme) and the evolved state (shown for all schemes).
In the `standard' scheme (blue lines), pairing instability caused by the cubic spline kernel and lack of mixing lead to an incorrect central solution in density (too high) and internal energy (too low).
The addition of the WC4 kernel (purple lines) prevents the formation of particle clumps at the centre and AC promotes fluid mixing but full AC without gravity limiter (green lines) leads to a numerical transport of internal energy outwards.
The gravity limiter treats this behaviour (brown lines).
The `new' scheme (red lines) with WC4 and AC and gravity limiter significantly improves the radial profiles in all physical quantities.}
  \label{fig:test_sphere}
\end{center}
\end{figure*}

Fig. \ref{fig:test_sphere} shows the results of the test problem at various times.
At first, the initial set-up (dotted lines) of the sphere is not yet in hydrostatic equilibrium and requires some time to settle.
Once hydrostatic equilibrium is reached around time $t=2.6$ (dashed lines for the `new' scheme) all hydrodynamical schemes must preserve the structure of the sphere and the radial profiles towards the final simulation time $t=7.6$.
Between all schemes, the thermal pressure profiles are indistinguishable balancing the gravitational pressure.
However, the composition of the thermal pressure $P=(\gamma-1)\rho{}u$ changes and the radial profiles for density $\rho$ and internal energy $u$ change significantly.
The `standard' scheme (blue lines) reaches the highest central density and also features lowest central internal energy and entropy.
However, the internal energy drops towards the centre and no stable solution is reached at all.
We suggest this behaviour to be a joint impact of pairing instability caused by the cubic spline kernel function and lack of internal energy mixing.
When introducing only AC, no gravitational limiter and no further SPH developments (green lines), the results marginally improve.
Internal energy still drops towards the centre but this time because too much AC was introduced.
In principle, AC leads to entropy cores but without the gravitational limiter over-mixing occurs and internal energy is transported from the centre to the outskirts along the pressure gradient.
The addition of the gravitational limiter (brown lines) improves the results significantly.
The divergence of profiles towards the centre is removed and a stable core of internal energy and entropy is reached.
Furthermore, no numerically induced transport of heat takes place.
Additionally, we perform a test run with introducing only the WC4 kernel and no further SPH improvements (purple lines).
In this run, the central density is lower than in the `standard' scheme, most probably because the clumping of particles in the centre and the occurance of the pairing instability is suppressed by the WC4 kernel in contrast to the run with the cubic spline.
This also leads to a plateau in internal energy in the centre of the sphere.
At last, we show the results of the entire `new' scheme (red lines) and find the results to remain stable with all additional modifications to WC4 kernel, AV and time-step limiter.
We find this run to give the most stable radial profiles in time.
This test confirms the importance of kernel functions immune to pairing instability and a proper implementation of AC and clearly shows the effects of particle clumping and over- and under-mixing in gravitationally virialized systems, which are important in cosmological simulations.

\subsection{Evrard collapse}

We consider the Evrard collapse \citep{evrard88} to study the SPH behaviour in the presence of dynamically important gravitational forces and collapse of gas.
We initialize a sphere of gas with mass $M=1$, radius $R=1$ and density profile of $\rho\sim{}r$ for $r<R$ and use vacuum boundary conditions and a gravitational softening length of $0.005$.
We do not use an external gravitational potential, dark matter particles or radiative cooling and thus the cloud only self-gravitates on the free-fall time-scale.
The gas is initially at rest and the thermal energy budget is orders of magnitude smaller than the gravitational binding energy.

\begin{figure*}
\begin{center}
  \includegraphics[width=0.95\textwidth]{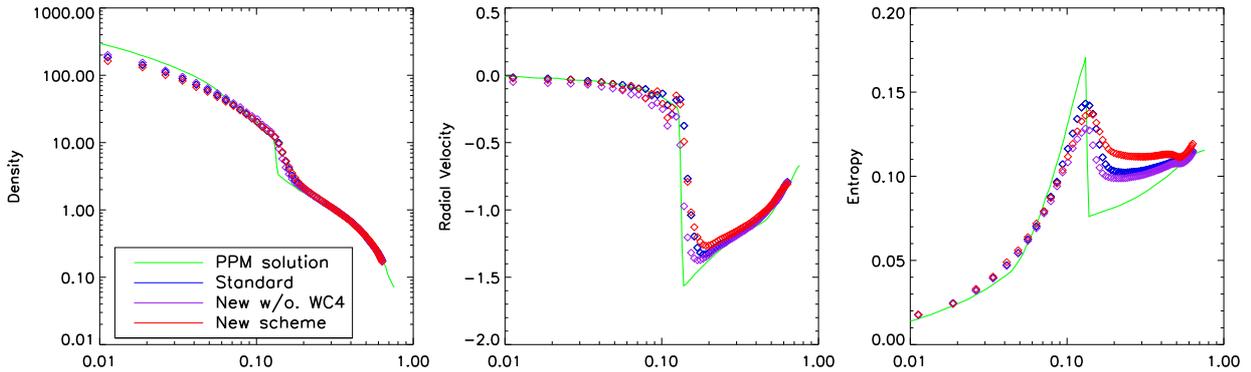}
 \caption{Evrard collapse.
We show radial profiles of density, velocity and entropy at time $t=0.8$ and compare to a reference piecewise parabolic grid computation (green lines).
In principle, all schemes (red, blue and purple lines) show similar characteristics, but they differ as follows.
The pre-shock entropy level is significantly higher in the `new' scheme (red lines), which we attribute to the WC4 kernel with a larger smoothing size.
A comparison run with the all improvements, but no WC4 kernel (purple lines), shows a lower pre-shock entropy level.}
 \label{fig:test_evrard}
\end{center}
\end{figure*}

Fig. \ref{fig:test_evrard} shows the results of the test problem at time $t=0.8$
We compare the SPH results to a reference solution similar to \cite{steinmetz93}.
In density (left panel) as well as in velocity (middle panel) all schemes show similar trends.
The general structure of the test is well reproduced; however, the shock front is slightly smoothed and broadened.
The solution in density deviates slightly from the reference solution at the centre.
The most striking differences can be seen in pre-shock entropy (right panel).
The `new' scheme (red line) produces higher levels of entropy compared to the `standard' scheme (blue line)
We investigate this trend and perform another test simulation (purple line) with the `new' scheme but the cubic spline kernel.
This run is closest to the pre-shock reference solution.

\subsection{Zel'dovich pancake}

We consider the Zel'dovich pancake \citep{zeldovich70} to study the SPH behaviour for the cosmological time integration with Hubble function $H(t)$ instead of time $t$.
This test describes the evolution of a sinusoidal cosmological perturbation in an expanding Einstein-de-Sitter universe.
After an initial linear growth phase, the one-dimensional perturbation collapses and several strong shocks develop.
Conveniently, the Zel'dovich pancake has an analytical solution describing the evolution well up to the collapse, which we used to create the initial conditions of the simulation.
The comoving position $x$ of an initially unperturbated coordinate $q$ at redshift $z$ is given by
\begin{equation} x(q,z) = q - \frac{1+z_c}{1+z}\frac{\sin(kq)}{k},\label{eq:zel}\end{equation}
where $k=2\pi/\lambda$ is the wavenumber of the perturbation with a wavelength of $\lambda$.
We numerically invert Eq. (\ref{eq:zel}) to obtain $q(x)$.
The peculiar velocity corresponding to the initial displacement is given by
\begin{equation}v_{\rm pec}(x,z) = -H_0 \frac{1+z_c}{(1+z)^{1/2}}\,\frac{\sin(kq)}{k},\end{equation}
and the comoving density is given by
\begin{equation}\rho(x,z) = \frac{\rho_0}{1- \frac{1+z_c}{1+z}\cos(kq)},\end{equation}
where $\rho_0$ is the critical density, $H_0$ the present-day Hubble constant and $z_c$ the redshift of collapse.
Furthermore, the temperature evolves adiabatically up to the collapse as
\begin{equation}T(x,z) = T_i \left[ \left(\frac{1+z}{1+z_i}\right)^3 \frac{\rho(x,z)}{\rho_0}\right]^{2/3},\end{equation}
where $z_i$ is the initial redshift.
We follow \cite{bryan95}, \cite{trac04} and \cite{springel10b} in our test set-up and choose $\lambda=64$ Mpc/h, $z_c=1$, $z_i=100$ and $T_i =
100\,{\rm K}$.
In a fully three-dimensional box, we set up 256$^3$ dark matter particles of equal masses as well as 256$^3$ gas particles of equal masses.

\begin{figure*}
\begin{center}
  \includegraphics[width=0.95\textwidth]{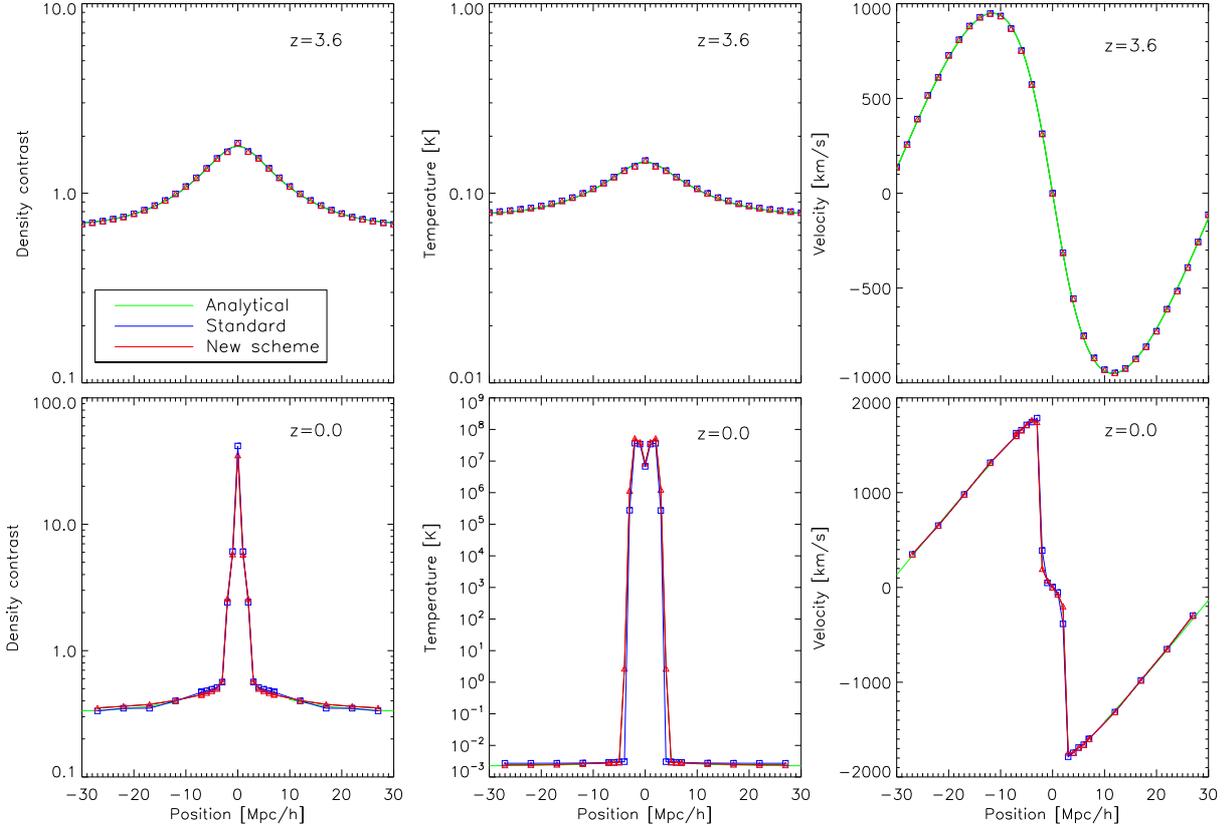}
  \caption{Zel'dovich pancake.
We show the evolution of density contrast (left column), temperature (middle column) and velocity (right column).
We show the state of the pancake at an intermediate redshift $z=3.6$, while it is still in the linear regime before the collapse and at the final redshift $z=0$, when it is evolved well into the non-linear regime.
Both SPH schemes agree well with the analytical solution during the linear growth phase.
At the final redshift, the `new' scheme resolves well the density contrast and yields a broader shock and a slightly smoother velocity.}
  \label{fig:test_zeldovich}
\end{center}
\end{figure*}

Fig. \ref{fig:test_zeldovich} shows the results of the test problem at two redshifts.
In the top row we show the evolution of the pancake before the collapse at redshift $z=3.6$, while it is still in the linear phase.
The `standard' (blue lines) and `new' (red lines) schemes give comparable results in density contrast (left column), temperature (middle column) and velocity (right column).
The simulated evolution agrees well with the analytical solution (green lines), which describes the linearized evolution of initial sinusoidal perturbation.
At this stage of the collapse the test is dominated by gravitational forces and hydrodynamical forces are negligible.
Therefore, we do not expect striking differences to arise between both schemes.
Both capture the linear growth and adiabatic evolution well.

In the bottom row we show the pancake at the final redshift $z=0$.
Again, we compare to the analytical solution in the regions outside the central shock.
In general, both schemes agree but we note the following differences.
The peak density contrast is marginally lower in the `new' scheme because of additional smoothing introduced by AC.
However, the evolution of density contrast in the low-density regions is described better by the `new' scheme and is resolved with less noise.
We recall the Sod shock tube (Section 3.1) and the Sedov blast wave (Section 3.2) problems, where we also find more accurately resolved density fields and lower peak densities.
Concerning temperature, we find the central shock to be slightly broader in the `new' scheme, which is caused by two effects.
Firstly, the higher amount of viscosity within the shock leads to an earlier heating of particles and thus broadens the shocks.
Secondly, the time-step limiting particle wake-up scheme captures highly active particles before they penetrate into inactive regions, which leads to a better fluid sampling and also shock broadening.
This can also be noticed in velocity, where the profile is slightly smoothed in the central region.
In summary, our `new' scheme gives reasonable results in this cosmological test problem and the differences between both schemes are very small at redshift $z=0$.
Therefore, our new implementation is ready to be applied to Idealized astrophysical problems.

%#########################################################################################################
%########################### Galaxy and galaxy cluster ###################################################
%#########################################################################################################

\section{Astrophysical applications}

We complete the evaluation of the performance and accuracy of the two different SPH implementations in idealized simulations of galaxy and galaxy cluster formation.

\subsection{Idealized galaxy formation}

In order to check if the `new' scheme is numerically stable when coupled with a simple effective description of the interstellar medium, we consider the formation of an isolated disk from a cooling gas cloud embedded within a rotating dark matter halo.
This idealized application also includes a the prescription for cooling, supernova feedback and star formation of \citep{springel03}.
We focus on the differences between both hydrodynamical schemes and therefore, we do not consider a cosmological environment or more advanced physical processes such as black holes, stellar evolution or metals.
Numerical comparison simulations \citep[see e.g.][]{scannapieco12} are a common tool to study the impact of numerical schemes and physical modules.

Within a computational domain of roughly 1 Mpc$^3$ we set up 4041345 particles resembling a Milky Way-like dark matter halo with a total mass of $1.8\cdot{}10^{12}$ M$_{\odot}$.
We include 4466429 gas particles with a total mass of $2.2\cdot{}10^{11}$ M$_{\odot}$, which corresponds to a baryon fraction of approximately 11 per cent.
The gravitational softening length is 450 pc.
Initially, the distribution of dark matter follows a Navarro-Frenk-White profile \citep{navarro97} and subsequently, we add the gaseous component similar to the set-up of the hydrostatic test (Section 4.1).
The only change is that here, we give the gas and dark matter a rotational velocity which peaks at 180 km s$^{-1}$.
Obviously, the initial hydrostatic equilibrium is broken by the onset of the gas cooling and we follow the evolution of the cloud for 10 Gyrs.

\begin{figure*}
\begin{center}
  \centerline{
  \includegraphics[width=0.475\textwidth]{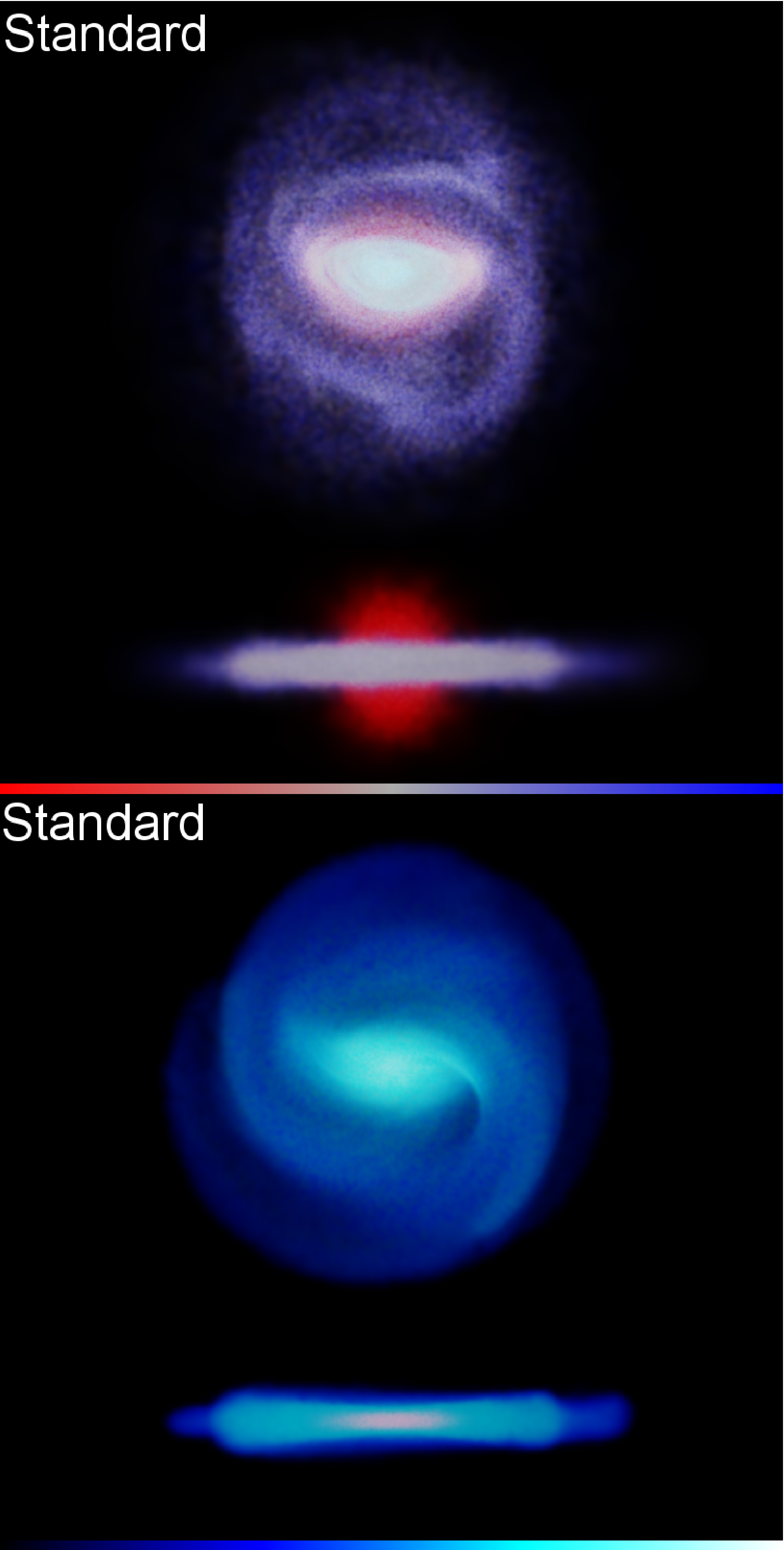}
  \includegraphics[width=0.475\textwidth]{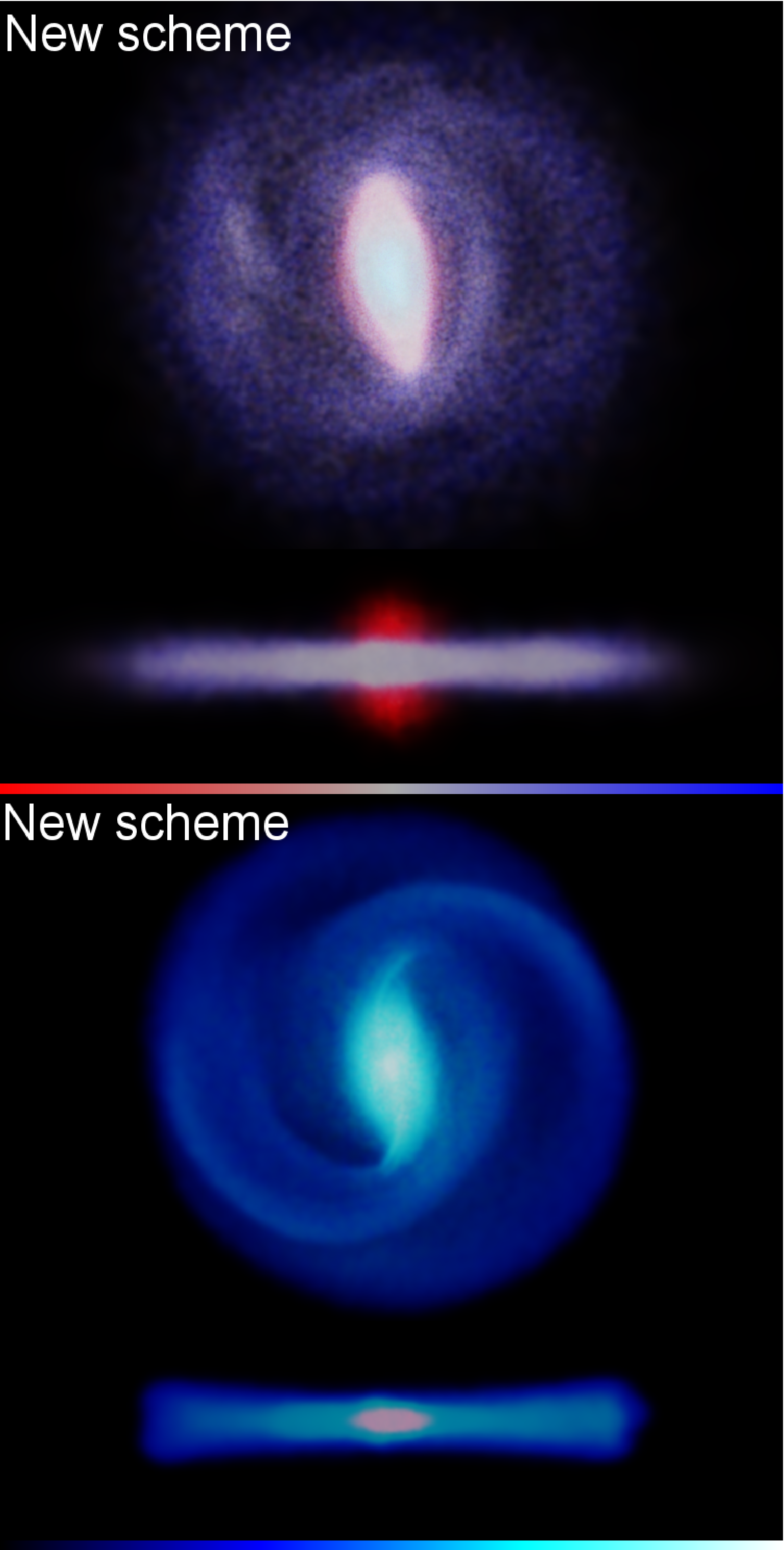}}
  \caption{Idealized galaxy formation.
We show the spatial distribution of stars, where the colours visualize the age of stars (top panels) and the spatial distribution of gas, where the colours visualize the star formation rate (bottom panels) at time $t=7.5$ Gyr.
In the `standard' scheme (left panels) the distribution of star formation is clumpy and the object appears bulgy.
In the `new' scheme (right panels) the distribution of star formation as well as the stellar component are more extended and pronounced in a disk-like structure.
Furthermore, the size of the bulge is smaller and the distribution of young stars is more extended.
We show a more quantitative comparison in Figs. \ref{fig:disk_stars} and \ref{fig:disk_gas}.}
  \label{fig:gal_projs}
\end{center}
\end{figure*}

Fig. \ref{fig:gal_projs} visualizes the spatial distribution of stars, where the colours represent the age of stars (top panels), and the spatial distribution of star forming gas, where the colours represent the star formation rate (bottom panels) at time $t=7.5$ Gyr.
We use the ray-tracing program SPLOTCH \citep{dolag08,jin10} to create the images and choose a linear colour bar for stellar age, where the red end of the colour bar corresponds to stars older than 3 Gyrs and the blue end to recently formed stars.
We visualize the star formation rate since it traces the (cold) gas within the disk and use a linear colour bar, which ranges from the minimum to the maximum value of star formation rate.
We choose identical plot settings for the `standard' scheme (left panels) and the `new' scheme (right panels), which show striking morphological differences as follows.

In the `standard' scheme, the galaxy shows a prominent bulge containing a large fraction of the stellar population.
The entire galaxy appears more spheroidal with a dominant bulge and the stellar disk is not well pronounced.
We find similar features in the distribution of star formation.
The gas disk is asymmetric and only shows little spiral structure in the face-on view.
In the edge-on projection the disk shows a rolling pin morphology.
Both disks are dominated by bulges, but in the `new' scheme the bulge is significantly less dominant.
The bulge contains a smaller fraction of the stellar population and might eventually be associated with an elliptical bar structure.
More as well as younger stars are present within the disk.
The gas disk is symmetric and shows a defined spiral structure.
At this stage of code testing it is difficult to track down the impact of individual code changes.
However, we assume the most significant differences are caused as follows.
In the `standard' scheme large amounts of AV might lead to dissipation of kinetic energy, loss of rotational support and numerical angular momentum transport.
Additionally, the mixing problem and its associated numerical surface tension tend to confine cold and dense gas blobs.
In the `new' scheme, significantly smaller amounts of AV are applied and rotational support can be provided.
Furthermore, the inclusion of AC promotes gas mixing between hot and cold phases.

\begin{figure*}
\begin{center}
  \includegraphics[width=0.95\textwidth]{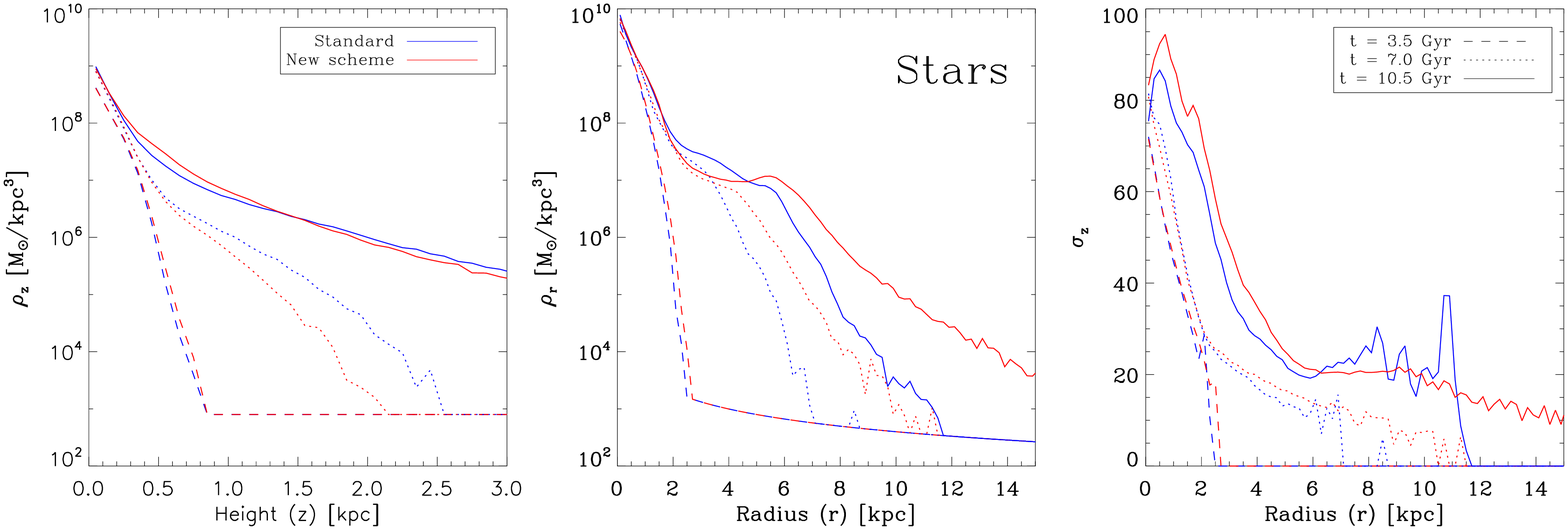}
  \caption{Idealized galaxy formation.
We show the vertical ($z$) density profile (left panel), the radial density profile (middle panel) and vertical velocity dispersion (right panel) of the stellar component.
For the radial plots we use cylindrical bins.
In the `new' scheme (red lines), the galactic disk is more defined, extended and colder.}
  \label{fig:disk_stars}
\end{center}
\end{figure*}

We continue with a more quantitative comparison of both schemes, which confirms our previous findings.
Fig. \ref{fig:disk_stars} shows density in vertical and radial direction ($\rho_\mathrm{z}$ and $\rho_\mathrm{r}$) of the stellar component as well as the associated vertical velocity dispersion ($\sigma_\mathrm{z}$) at times 2.5 (dashed lines), 5.0 (dotted lines) and 7.5 Gyrs (solid lines).
As seen in $\rho_\mathrm{z}$ (left panel), the `new' scheme produces a thinner stellar disk and as seen in $\rho_\mathrm{r}$ (middle panel) the disk also extends to significantly larger radii.
This trend is confirmed by $\sigma_\mathrm{z}$ (right panel), which for the `standard' scheme truncates at smaller radii than for the `new' scheme.

\begin{figure*}
\begin{center}
  \includegraphics[width=0.95\textwidth]{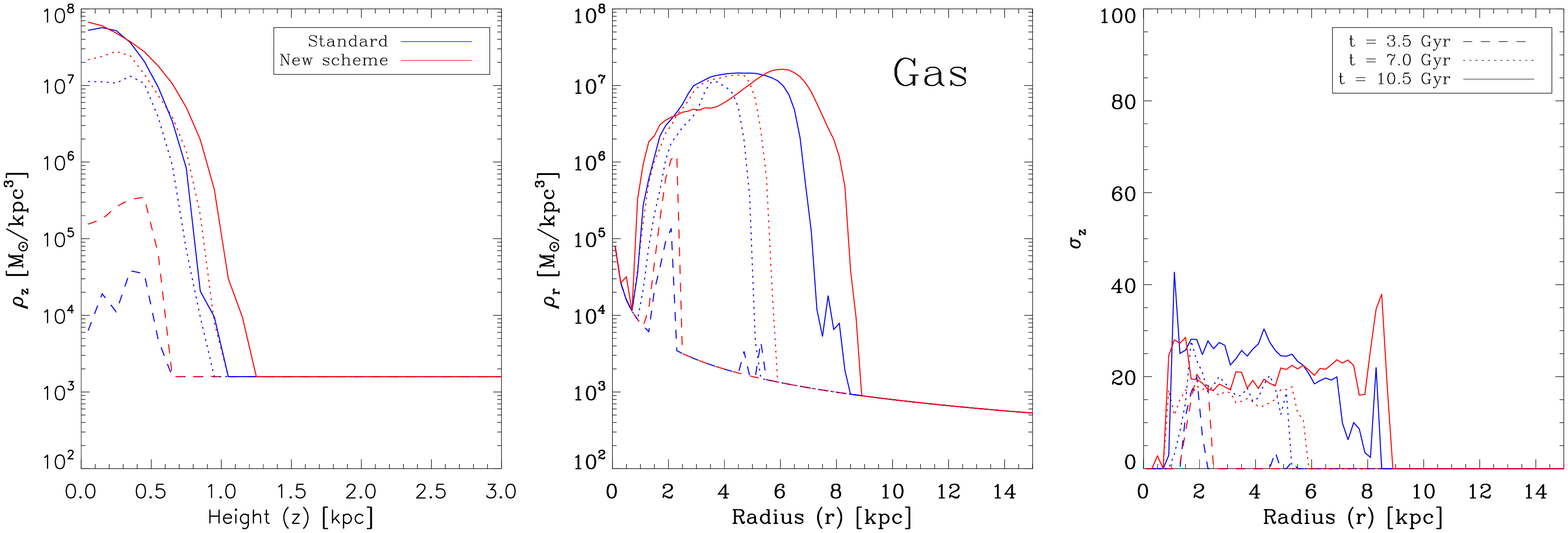}
  \caption{Idealized galaxy formation.
We show the vertical ($z$) density profile (left panel), the radial density profile (middle panel) and vertical velocity dispersion (right panel) of the cold gas component.
For the radial plots we use cylindrical bins.
We use a temperature criterion of $T<10^{5}$K to select the cold gas.
In the `new' scheme (red lines), the galactic disk is more defined, extended and colder.}
  \label{fig:disk_gas}
\end{center}
\end{figure*}

Fig. \ref{fig:disk_gas} shows vertical and radial profiles of gas density ($\rho_\mathrm{z}$ and $\rho_\mathrm{r}$) as well as the vertical gas velocity dispersion ($\sigma_\mathrm{z}$) of the cold gas at times 2.5 (dashed lines), 5.0 (dotted lines) and 7.5 Gyrs (solid lines).
We employ a temperature criterion of $T<10^{5}$K to distinguish between cold and hot gas.
$\rho_\mathrm{r}$ decreases towards the centre since the gas within the bulge is hot and exceeds our temperature threshold.
In the `new' scheme, the distribution of cold gas is slightly more extended in vertical as well as radial direction.
However, $\sigma_\mathrm{z}$ indicates a colder gas disk.
Most probably, these features are a result of less numerically induced AV, angular momentum transport and depression of rotational support.
Furthermore, the inclusion of AC allows mixing between gas phases and promotes dissociation of cold structures.

\subsection{Santa Barbara Cluster}

We carried out the Santa Barbara galaxy cluster \citep{frenk99}, which is a common reference simulation for cosmological hydrodynamical simulation codes.
Although no analytic solutions exists, the cluster has been simulated with a large variety of different codes.
The simulation describes the formation of a massive dark matter halo, with virial mass of $1.2\cdot{}10^{15}$ M$_{\odot}$ and virial radius of 2.8 Mpc.
It is evolved in an Einstein-de Sitter cold dark matter cosmology with parameters of $\Omega_{M}=1.0$, $\Omega_{\Lambda}=0.0$, and $H_{0}=50$ km s$^{-1}$ Mpc$^{-1}$.
We choose an initial redshift of $z=20$ with a perturbed distribution of $256^3$ dark matter particles and $256^3$ gas particles, each of equal masses \citep[see also][for a detailed description of the initial conditions]{frenk99}, and follow the formation until redshift $z=0$.

\begin{figure*}
\begin{center}
  \includegraphics[width=0.95\textwidth]{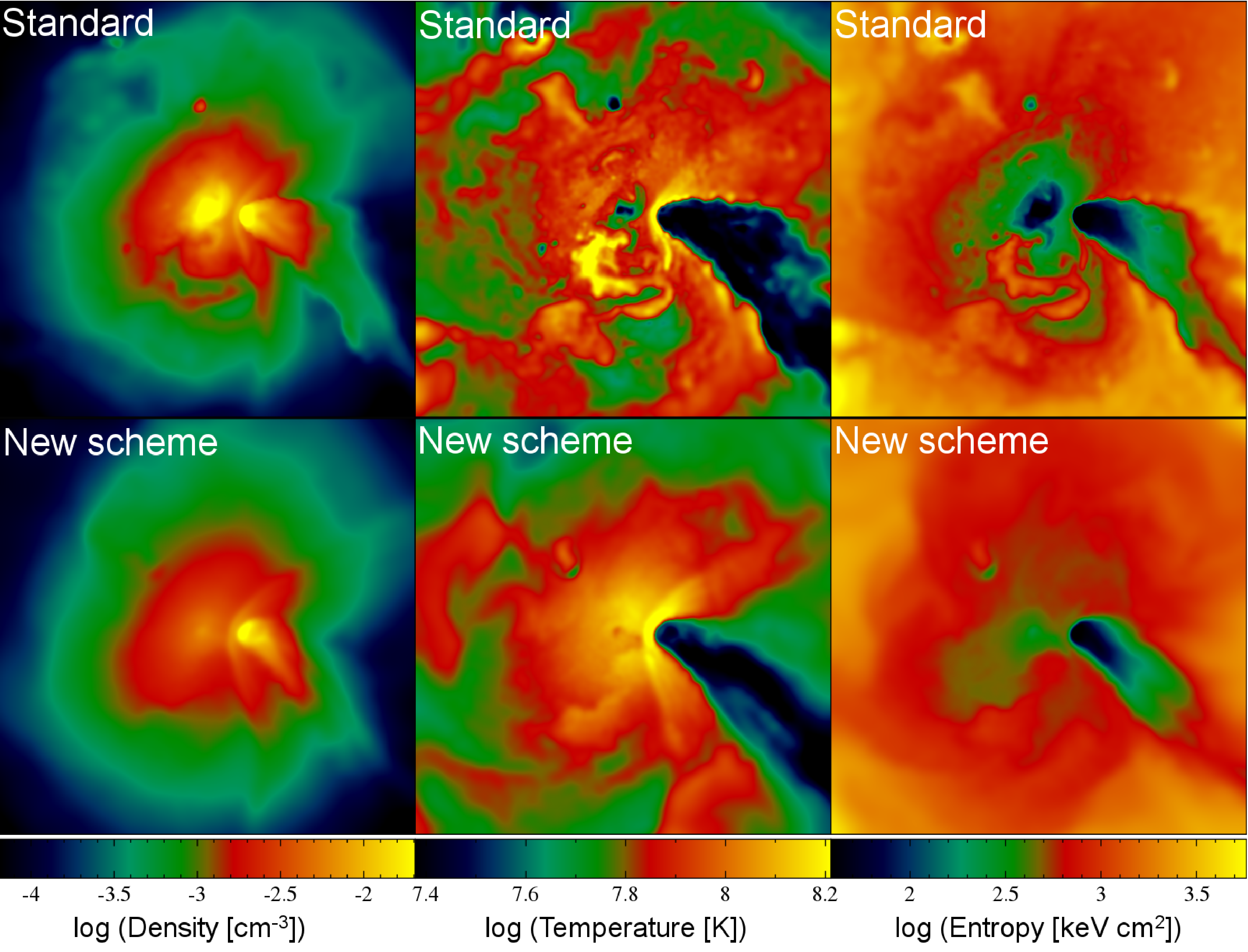}
  \caption{Santa Barbara Cluster.
In boxes with 1 Mpc/h side length, we show this slices of gas density (left panels), temperature (middle panels) and entropy (right panels) at redshift $z=0$ for the `standard' scheme (top row) and the `new' scheme (bottom row).
In the `new' scheme significantly less dense and cold gas blobs are present, as AC promotes fluid mixing and blob dissociation.
This promotes a smoother distribution of higher temperatures and entropies at the halo centre, which reduces the ICM complexity.}
  \label{fig:sb_maps}
\end{center}
\end{figure*}

Fig. \ref{fig:sb_maps} shows thin slices of gas density (left panels), temperature (middle panels) and entropy (right panels) defining the thermodynamical state of the hot intracluster medium (ICM) at redshift $z=0$.
For the definition of entropy we use $S=T/{n_{e}}^{2/3}$, which is commonly used in X-ray studies of the ICM \citep[e.g.][]{kravtsov12}.
From the maps we note the following interesting features.

The gas density (left panels) tends to be smoother in the `new' scheme, which is mostly due to the effect of AC, which introduces entropy mixing among neighbouring gas particles.
In contrast, the `standard' scheme produces a clumpy distribution of gas with gas inhomogeneities associated to stripping from merging haloes and cold blobs.
These structures are persistent in the hot ICM mainly due to the lack of mixing.
In turn, these 'features' of the `standard' scheme prevent an efficient action of hydrodynamical instabilities such as Rayleigh-Taylor and Kelvin-Helmholtz instabilities that are spuriously inhibited.
Quite remarkably, the clumps are much less evident in the `new' scheme, which also produces a lower value for the central gas density.
In the temperatures slices (middle panels) it becomes clear that gas clumps in the `standard' scheme correspond to objects of low temperature.
As expected, the effect of introducing AC is the reduction of the degree of ICM thermal complexity.  

However, in the `new' scheme the bow-shock, which is induced by the infall of a large substructure onto the main halo is better defined than in the `standard' scheme.
The bow-shock is located on the right hand side of the main halo centre (see Fig. \ref{fig:sb_maps}).
In fact, the WC4 kernel of the `new' scheme captures the entropy discontinuity associated to the shock better.
Most importantly, we note from the entropy maps (right panels) that the entropy level in the innermost region of the clusters increases in the `new' scheme.

\begin{figure*}
\begin{center}
  \includegraphics[width=0.95\textwidth]{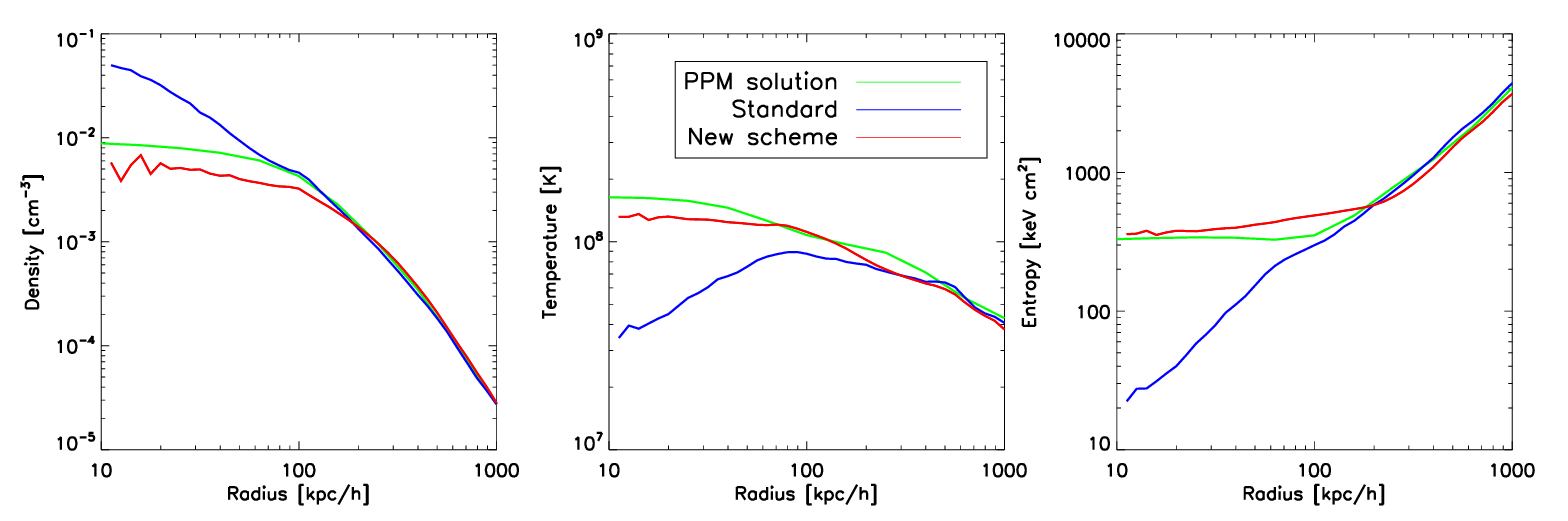}
  \caption{Santa Barbara Cluster. 
We show radial profiles of gas density (left panel), temperature (middle panel) and entropy (right panel).
In each panel, we compare the results of the `standard' (blue lines) and the `new' (red lines) scheme to a reference solution (black lines) obtained with a piecewise parabolic grid computation with the MASCLET code \citep[see][]{quilis04}.
The `standard' scheme does not produce an entropy core but a diverging profile towards the halo centre.
In the `new' scheme an entropy core as well as stable temperature and density profiles are reached, which are all in good agreement to the grid code computation.}
  \label{fig:sb_profs}
\end{center}
\end{figure*}

Fig. \ref{fig:sb_profs} shows radial profiles of gas density, temperature and entropy in the same units as shown in Fig. \ref{fig:sb_maps} at redshift $z=0$ for both SPH schemes.
Furthermore, for a comparison, we also include results obtained with the MASCLET grid code \citep[see][]{quilis04}.
Both schemes produce quite consistent results at relatively large radii ($>200$kpc/h) but show striking differences in the innermost regions.
The `new' scheme predicts a flatter central gas density profile, also with no evidence for the inversion of the temperature gradient produced by the `standard' scheme.
Density and temperature profiles for the `new' scheme combine to produce a flat entropy core, which extends out to $\approx{}100$kpc/h.
In the `standard' scheme, the persistence of cold and dense clumps in the cluster atmosphere causes their low-entropy gas to sink to the centre of the cluster, thereby causing the continuous decrease of entropy.
In the `new' scheme, AC promotes mixing of gas phases and helps to dissolve low-entropy gas blobs within the hot ICM atmosphere and causes a higher entropy level to be established.

The results for the `new' scheme are remarkably similar to those of the MASCLET code and, more in general, reported by \cite{frenk99,vazza11,power14} for Eulerian codes.
We point out that such a close agreement has been obtained without any tuning aimed at producing the entropy core predicted by Eulerian codes in cosmological simulations for the formation of galaxy clusters.
The choice of parameters for the `new' SPH scheme was only aimed at preventing the limitations of `standard' SPH in terms of the description of discontinuities and efficiency to capture hydrodynamical instabilities.
Note also that these results for the Santa Barbara cluster are in qualitative agreement with those obtained in the hydrostatic sphere (see Section 4.1).
The behaviour of both schemes are in the same direction, even if they are less evident than in this Santa Barbara cluster test, due to the lack of the hierarchical process of structure formation within a cosmological environment.

Additionally, we analyze the simulation also at redshifts $z>0$.
In general, the profiles of the high-redshift haloes show the same behaviour as their low-redshift counterparts, provided that we choose quiet and virialized objects.
Objects, which host dynamically important shocks or undergo merger events show altered radial profiles because the timing of the mergers depends slightly on the simulation scheme.
Therefore, a sensible comparison of the entire redshift-evolution and behaviour of both schemes during these violent phases of structure formation is not possible and requires controlled experiments.

%#########################################################################################################
%########################### Summary #####################################################################
%#########################################################################################################

\section{Summary and conclusions}

In this paper we presented a novel implementation of the SPH scheme in the GADGET code, which provides improved accuracy for simulations of galaxies and large-scale cosmic structures. 
Since the first development of SPH great advancements have been made to improve the reliability and stability of this hydrodynamical scheme and, in particular, much effort has been spent in a proper treatment of discontinuities. 
We implemented and improved several of these modifications of SPH into the developer version of GADGET-3, and tested them against a number of standard hydrodynamical problems, as well as first simple astrophysical applications.
The main modifications (see also Table \ref{tab:comp}) of this `new' scheme, when compared to the `standard' \citep[see e.g.][]{price12a} formulation of SPH, can be summarised as follows.

\begin{table*}
\begin{center}
  \begin{tabular}{@{}lccccccl}
    \hline\hline
    & AV & AC & WakeUp & Grav. & Cosmo & Radiat. & Main result \\\hline\hline
    Sod shock tube & \checkmark & \checkmark & - & - & - & - & Smooth density field / Sharp shock front / No pressure-blip\\
    Sedov blast wave & \checkmark & \checkmark & \checkmark & - & - & - & Sharp shock front / Central temperature profile\\
    Keplerian ring & \checkmark (imp.) & - & - & - & - & - & Stability of ring / No angular momentum transport\\
    Cold blob test & \checkmark & \checkmark & - & - & - & - & No surface tension / Mixing of gas phases\\
    KH instability & \checkmark (imp.) & \checkmark & - & - & - & - & Growth of perturbation / Mixing of gas phases\\
    Turbulent boxes & \checkmark (imp.) & - & - & - & - & - & Steady-state spectrum / Preservation of kinetic energy\\
    Hydrostatic test & - & \checkmark (imp.) & - & \checkmark & - & - & Stability in radial profiles of density and entropy\\
    Evrard collapse & \checkmark (imp.) & \checkmark (imp.) & \checkmark & \checkmark & - & - & Radial profiles in density and entropy\\
    Zel'dovich pancake & \checkmark & \checkmark (imp.) & \checkmark & \checkmark & \checkmark & - & Smooth density field / Sharp shock front\\
    Idealized galaxy & \checkmark (imp.) & \checkmark (imp.) & - & \checkmark & - & \checkmark & Extended stellar and gas disks\\    
    SB cluster & \checkmark (imp.) & \checkmark (imp.) & \checkmark & \checkmark & \checkmark & - & Formation of entropy core / Dissociation of cold blobs\\\hline\hline
  \end{tabular}
  \caption{Overview of our test problems.
For each test we note the relative importance of a standard method (\checkmark) or an improved method (\checkmark (imp.)) of artificial viscosity (AV), artificial conductivity (AC), time-step limiter (WakeUp) and the physical processes involved beyond pure hydrodynamics such as gravity (Grav.), cosmological time integration (Cosmo), radiative cooling, star formation and supernova feedback (Radiat.).}
  \label{tab:tests}
\end{center}
\end{table*}

\begin{itemize}
\item Artificial viscosity (AV) is introduced for a proper description of shocks.  It prevents particle interpenetration into unshocked regions and provides a regularisation of the particle field, which supports a proper sampling of the fluid.
First spatially constant low-order formulations of AV introduce viscosity not only at shocks, but also within unshocked regions and shearing flows, thereby leading to an overly viscous behavior and a too fast dissipation of kinetic energy.
Most commonly, the so called Balsara switch \citep{balsara95} is used to reduce viscosity in shear flows, while further attempts were made to reduce AV where it is unwanted \citep{morris97,dolag05}.  Recently, modern formulations of AV \citep{cullen10,hu14} improved greatly on a correct detection of shocks and use high-order gradient estimators to calculate divergence and curl of velocity from the full velocity gradient matrix instead of the classical SPH estimators.
This allows shear flow limiters, such as the Balsara one, to work more accurately and suppress AV outside shocks and in shearing flows.
In this way, kinetic energy is better preserved, thus helping simulating turbulent flows or hydrodynamical instabilities with higher accuracy. 
In our `new' scheme, we compute the velocity gradients from the full velocity gradient matrix instead of low-order classic kernel derivatives.

\item Artificial conductivity (AC) is introduced to provide a proper fluid description at contact discontinuities.
In fact, in the density-entropy formulation a spurious surface tension arises at discontinuities, which also suppresses the formation of instabilities and prevents mixing of different fluid phases.
Lately, AC \citep[see e.g.][]{price08} or pressure-entropy formulations \citep{hopkins13,saitoh13} were proposed to overcome these issues.
AC is applied at contact discontinuities and promotes the transport of heat between particles.
However, in the presence of gravitationally induced pressure or temperature gradients, common AC schemes might lead to unwanted transport opposite to the gravitational force.
Therefore, numerical limiters are necessary to be included in AC.
In our `new' scheme, we include locally adaptive AC to transport heat and treat contact discontinuities in SPH and we limit the amount of AC by correcting for gravitationally induced pressure gradients. While we demonstrated that our AC model is quite efficient at reducing such a surface tension, admittely a small residual effect is still present, and could potentially impact the long-term stability by over-diffusion.

\item As for the choice of the interpolating kernel, the commonly employed cubic spline function has been shown to become easily unstable, which leads to spurious pairs of particles, incorrect gradient estimators and, in general, a poor fluid sampling.
Therefore, a change of the kernel function is highly recommended, where commonly the Wendland kernels \citep{dehnen12} are now used.
In our `new' scheme, we employ the Wendland $C^{4}$ kernel function with 200 neighbours instead of a cubic spline with 64 neighbours.
We calculate the density in a classic fashion from the mass distribution of particles and also compute the hydrodynamical forces with the density-entropy formulation.

\item At last, within supersonic shocks highly dynamical and computationally active particles can penetrate into regions containing computationally inactive particles causing distortions in the fluid sampling and incorrect hydrodynamical solutions.
In our `new' scheme we use a particle wake-up time-step limiting scheme \citep[see][]{saitoh09,pakmor12} as a solution, so as to shorten the time-steps whenever necessary and allow particles to become active earlier.
\end{itemize}

To highlight the improvements associated to this advanced SPH implementation, we investigate both the new and the original scheme in a variety of hydrodynamical standard tests, with and without gravity.
Furthermore, we study the behavior in the cosmological problem of the formation of galaxy cluster and enable simple prescriptions for radiative cooling, supernova feedback and star formation for a test simulation of an isolated rotating disk galaxy.
Table \ref{tab:tests} presents an overview of our test problems and shows if SPH modules are \textit{important} in a standard (\checkmark) or improved (\checkmark(imp.)) with respect to GADGET-SPH without our modifications.
Furthermore, we list the probed physical features of each test.

The inclusion of AC in SPH changes the thermodynamical evolution of density, internal energy and pressure.
Additionally, physical conduction \citep[see e.g.][]{arth14} is also sometimes employed in cosmological SPH simulations to promote (an)isotropic heat transport, which also helps to overcome the limitations of `standard' GADGET-SPH.
The joint effect of artificial conduction, introduced for purely numerical reasons, and of physical conduction awaits further investigations.  

In summary, the `new' GADGET-SPH scheme presented here performs better than the `standard' one in every single of our test simulations.
Therefore, it provides a much improved numerical description for weakly collisionless plasmas in cosmological simulations down to galactic scales.
We base our future simulations of galaxies and large-scale cosmic structures on this updated formulation of SPH.
We will also carry out detailed studies of galactic magnetic fields \citep[see][]{beck13,beck14} and the ICM with this advanced method.  
In view of these applications, it is important to verify how this new SPH implementation performs when compared to other variants of SPH and to Eulerian codes.
To this purpose, this SPH implementation participated to the nIFTy cosmology comparison project \citep{sembolini15}, which compares the performances of different hydrodynamical codes in cosmological re-simulations of galaxy clusters.
In that comparison project, our code is shown to agree very well to both Eulerian codes and modern SPH implementations on the radial profiles of gas density, temperature and entropy.
Given the improvements in the description of hydrodynamics provided by the new SPH implementation presented here, we regard it as the core of an efficient code for modern simulations of cosmic structure formation.

%#########################################################################################################
%########################### Acknowledgments #############################################################
%#########################################################################################################

\section*{Acknowledgments}
We thank the anonymous referee for extensive and detailed comments, which helped to improve the quality and presentation of the paper.
We thank Volker Springel for always granting access to the developer version of GADGET.
We thank Elena Rasia, Veronica Biffi, Milena Valentini, Thorsten Naab, David Hubber and David Schlachtberger for important discussions about numerical topics and SPH.
We thank Vicent Quilis for several grid code comparison simulations with MASCLET \citep{quilis04}.
We thank Margarita Petkova and the Computational Centre for Particle and Astrophysics (C2PAP) for technical support.
AMB thanks the Osservatorio Astronomico di Trieste for its hospitality during several stays.
AMB and several authors are part of the MAGNETICUM\footnote[3]{http://www.magneticum.org} simulation core team.
Rendered plots are created using the SPLASH software written by Daniel Price \citep[see][]{price07}.
AMB, AA, R-SR, AFT and KD are using SuperMuc of the Leibniz-Rechenzentrum via project 'pr86re'.
AMB, AFT and KD are supported by the DFG Research Unit 1254 'Magnetisation of Interstellar and Intergalactic Media' and by the DFG Cluster of Excellence 'Origin and Structure of the Universe'.
AFT and KD are supported by the DFG Sonderforschungsbereich TRR 33 'The Dark Universe'.
JMFD is supported by the EU FP7 Marie Curie program 'People'.
GM, SB and SP are supported by the 'Consorzio per la Fisica di Trieste', the PRIN-INAF12 grant 'The Universe in a Box: Multi-scale Simulations of Cosmic Structures', the PRIN-MIUR 01278X4FL grant 'Evolution of Cosmic Baryons', and the INDARK INFN grant.
SP is supported by the 'Spanish Ministerio de Ciencia e Innovaci\'on (MICINN) with the grants AYA2010-21322-C03-02 and CONSOLIDER2007-00050.

%#########################################################################################################
%########################### Final stuff #####################################################################
%#########################################################################################################

\bibliographystyle{mnras}
\bibliography{sph.bib}

\begin{thebibliography}{}
\makeatletter
\relax
\def\mn@urlcharsother{\let\do\@makeother \do\$\do\&\do\#\do\^\do\_\do\%\do\~}
\def\mn@doi{\begingroup\mn@urlcharsother \@ifnextchar [ {\mn@doi@}
  {\mn@doi@[]}}
\def\mn@doi@[#1]#2{\def\@tempa{#1}\ifx\@tempa\@empty \href
  {http://dx.doi.org/#2} {doi:#2}\else \href {http://dx.doi.org/#2} {#1}\fi
  \endgroup}
\def\mn@eprint#1#2{\mn@eprint@#1:#2::\@nil}
\def\mn@eprint@arXiv#1{\href {http://arxiv.org/abs/#1} {{\tt arXiv:#1}}}
\def\mn@eprint@dblp#1{\href {http://dblp.uni-trier.de/rec/bibtex/#1.xml}
  {dblp:#1}}
\def\mn@eprint@#1:#2:#3:#4\@nil{\def\@tempa {#1}\def\@tempb {#2}\def\@tempc
  {#3}\ifx \@tempc \@empty \let \@tempc \@tempb \let \@tempb \@tempa \fi \ifx
  \@tempb \@empty \def\@tempb {arXiv}\fi \@ifundefined
  {mn@eprint@\@tempb}{\@tempb:\@tempc}{\expandafter \expandafter \csname
  mn@eprint@\@tempb\endcsname \expandafter{\@tempc}}}

\bibitem[\protect\citeauthoryear{{Agertz} et~al.,}{{Agertz}
  et~al.}{2007}]{agertz07}
{Agertz} O.,  et~al., 2007, \mn@doi [\mnras]
  {10.1111/j.1365-2966.2007.12183.x}, \href
  {http://adsabs.harvard.edu/abs/2007MNRAS.380..963A} {380, 963}

\bibitem[\protect\citeauthoryear{{Arth}, {Dolag}, {Beck}, {Petkova}  \&
  {Lesch}}{{Arth} et~al.}{2014}]{arth14}
{Arth} A.,  {Dolag} K.,  {Beck} A.~M.,  {Petkova} M.,   {Lesch} H.,  2014,
  preprint, \href {http://adsabs.harvard.edu/abs/2014arXiv1412.6533A} {}
  (\mn@eprint {arXiv} {1412.6533})

\bibitem[\protect\citeauthoryear{{Balsara}}{{Balsara}}{1995}]{balsara95}
{Balsara} D.~S.,  1995, \mn@doi [Journal of Computational Physics]
  {10.1016/S0021-9991(95)90221-X}, \href
  {http://adsabs.harvard.edu/abs/1995JCoPh.121..357B} {121, 357}

\bibitem[\protect\citeauthoryear{{Bauer} \& {Springel}}{{Bauer} \&
  {Springel}}{2012}]{bauer12}
{Bauer} A.,  {Springel} V.,  2012, \mn@doi [\mnras]
  {10.1111/j.1365-2966.2012.21058.x}, \href
  {http://adsabs.harvard.edu/abs/2012MNRAS.423.2558B} {423, 2558}

\bibitem[\protect\citeauthoryear{{Beck}, {Dolag}, {Lesch}  \&
  {Kronberg}}{{Beck} et~al.}{2013}]{beck13}
{Beck} A.~M.,  {Dolag} K.,  {Lesch} H.,   {Kronberg} P.~P.,  2013, \mn@doi
  [\mnras] {10.1093/mnras/stt1549}, \href
  {http://adsabs.harvard.edu/abs/2013MNRAS.435.3575B} {435, 3575}

\bibitem[\protect\citeauthoryear{{Beck}, {Beck}, {Beck}, {Dolag}, {Strong}  \&
  {Nielaba}}{{Beck} et~al.}{2014}]{beck14}
{Beck} M.~C.,  {Beck} A.~M.,  {Beck} R.,  {Dolag} K.,  {Strong} A.~W.,
  {Nielaba} P.,  2014, preprint, \href
  {http://adsabs.harvard.edu/abs/2014arXiv1409.5120B} {} (\mn@eprint {arXiv}
  {1409.5120})

\bibitem[\protect\citeauthoryear{{Biffi} \& {Valdarnini}}{{Biffi} \&
  {Valdarnini}}{2015}]{biffi15}
{Biffi} V.,  {Valdarnini} R.,  2015, \mn@doi [\mnras] {10.1093/mnras/stu2278},
  \href {http://adsabs.harvard.edu/abs/2015MNRAS.446.2802B} {446, 2802}

\bibitem[\protect\citeauthoryear{{Brunetti} \& {Lazarian}}{{Brunetti} \&
  {Lazarian}}{2007}]{brunetti07}
{Brunetti} G.,  {Lazarian} A.,  2007, \mn@doi [\mnras]
  {10.1111/j.1365-2966.2007.11771.x}, \href
  {http://adsabs.harvard.edu/abs/2007MNRAS.378..245B} {378, 245}

\bibitem[\protect\citeauthoryear{{Bryan}, {Norman}, {Stone}, {Cen}  \&
  {Ostriker}}{{Bryan} et~al.}{1995}]{bryan95}
{Bryan} G.~L.,  {Norman} M.~L.,  {Stone} J.~M.,  {Cen} R.,   {Ostriker} J.~P.,
  1995, \mn@doi [Computer Physics Communications]
  {10.1016/0010-4655(94)00191-4}, \href
  {http://adsabs.harvard.edu/abs/1995CoPhC..89..149B} {89, 149}

\bibitem[\protect\citeauthoryear{{Bryan} et~al.,}{{Bryan}
  et~al.}{2014}]{bryan14}
{Bryan} G.~L.,  et~al., 2014, \mn@doi [\apjs] {10.1088/0067-0049/211/2/19},
  \href {http://adsabs.harvard.edu/abs/2014ApJS..211...19B} {211, 19}

\bibitem[\protect\citeauthoryear{{Cartwright}, {Stamatellos}  \&
  {Whitworth}}{{Cartwright} et~al.}{2009}]{cartwright09}
{Cartwright} A.,  {Stamatellos} D.,   {Whitworth} A.~P.,  2009, \mn@doi
  [\mnras] {10.1111/j.1365-2966.2009.14720.x}, \href
  {http://adsabs.harvard.edu/abs/2009MNRAS.395.2373C} {395, 2373}

\bibitem[\protect\citeauthoryear{{Cha}, {Inutsuka}  \& {Nayakshin}}{{Cha}
  et~al.}{2010}]{cha10}
{Cha} S.-H.,  {Inutsuka} S.-I.,   {Nayakshin} S.,  2010, \mn@doi [\mnras]
  {10.1111/j.1365-2966.2010.16200.x}, \href
  {http://adsabs.harvard.edu/abs/2010MNRAS.403.1165C} {403, 1165}

\bibitem[\protect\citeauthoryear{{Cui}, {Liu}, {Yang}, {Wang}, {Feng}  \&
  {Springel}}{{Cui} et~al.}{2008}]{cui08}
{Cui} W.,  {Liu} L.,  {Yang} X.,  {Wang} Y.,  {Feng} L.,   {Springel} V.,
  2008, \mn@doi [\apj] {10.1086/592079}, \href
  {http://adsabs.harvard.edu/abs/2008ApJ...687..738C} {687, 738}

\bibitem[\protect\citeauthoryear{{Cullen} \& {Dehnen}}{{Cullen} \&
  {Dehnen}}{2010}]{cullen10}
{Cullen} L.,  {Dehnen} W.,  2010, \mn@doi [\mnras]
  {10.1111/j.1365-2966.2010.17158.x}, \href
  {http://adsabs.harvard.edu/abs/2010MNRAS.408..669C} {408, 669}

\bibitem[\protect\citeauthoryear{{Daubechies}}{{Daubechies}}{1992}]{daubechies92}
{Daubechies} I.,  ed. 1992, {Ten lectures on wavelets}

\bibitem[\protect\citeauthoryear{{Dehnen} \& {Aly}}{{Dehnen} \&
  {Aly}}{2012}]{dehnen12}
{Dehnen} W.,  {Aly} H.,  2012, \mn@doi [\mnras]
  {10.1111/j.1365-2966.2012.21439.x}, \href
  {http://adsabs.harvard.edu/abs/2012MNRAS.425.1068D} {425, 1068}

\bibitem[\protect\citeauthoryear{{Diehl}, {Rockefeller}, {Fryer}, {Riethmiller}
   \& {Statler}}{{Diehl} et~al.}{2012}]{diehl12}
{Diehl} S.,  {Rockefeller} G.,  {Fryer} C.~L.,  {Riethmiller} D.,   {Statler}
  T.~S.,  2012, preprint, \href
  {http://adsabs.harvard.edu/abs/2012arXiv1211.0525D} {} (\mn@eprint {arXiv}
  {1211.0525})

\bibitem[\protect\citeauthoryear{{Dolag}, {Vazza}, {Brunetti}  \&
  {Tormen}}{{Dolag} et~al.}{2005}]{dolag05}
{Dolag} K.,  {Vazza} F.,  {Brunetti} G.,   {Tormen} G.,  2005, \mn@doi [\mnras]
  {10.1111/j.1365-2966.2005.09630.x}, \href
  {http://adsabs.harvard.edu/abs/2005MNRAS.364..753D} {364, 753}

\bibitem[\protect\citeauthoryear{{Dolag}, {Reinecke}, {Gheller}  \&
  {Imboden}}{{Dolag} et~al.}{2008}]{dolag08}
{Dolag} K.,  {Reinecke} M.,  {Gheller} C.,   {Imboden} S.,  2008, \mn@doi [New
  Journal of Physics] {10.1088/1367-2630/10/12/125006}, \href
  {http://adsabs.harvard.edu/abs/2008NJPh...10l5006D} {10, 125006}

\bibitem[\protect\citeauthoryear{{Durier} \& {Dalla Vecchia}}{{Durier} \&
  {Dalla Vecchia}}{2012}]{durier12}
{Durier} F.,  {Dalla Vecchia} C.,  2012, \mn@doi [\mnras]
  {10.1111/j.1365-2966.2011.19712.x}, \href
  {http://adsabs.harvard.edu/abs/2012MNRAS.419..465D} {419, 465}

\bibitem[\protect\citeauthoryear{{Evrard}}{{Evrard}}{1988}]{evrard88}
{Evrard} A.~E.,  1988, \mnras, \href
  {http://adsabs.harvard.edu/abs/1988MNRAS.235..911E} {235, 911}

\bibitem[\protect\citeauthoryear{{Frenk} et~al.,}{{Frenk}
  et~al.}{1999}]{frenk99}
{Frenk} C.~S.,  et~al., 1999, \mn@doi [\apj] {10.1086/307908}, \href
  {http://adsabs.harvard.edu/abs/1999ApJ...525..554F} {525, 554}

\bibitem[\protect\citeauthoryear{{Gingold} \& {Monaghan}}{{Gingold} \&
  {Monaghan}}{1977}]{gingold77}
{Gingold} R.~A.,  {Monaghan} J.~J.,  1977, \mnras, \href
  {http://adsabs.harvard.edu/abs/1977MNRAS.181..375G} {181, 375}

\bibitem[\protect\citeauthoryear{{Hernquist} \& {Katz}}{{Hernquist} \&
  {Katz}}{1989}]{hernquist89}
{Hernquist} L.,  {Katz} N.,  1989, \mn@doi [\apjs] {10.1086/191344}, \href
  {http://adsabs.harvard.edu/abs/1989ApJS...70..419H} {70, 419}

\bibitem[\protect\citeauthoryear{{Hockney} \& {Eastwood}}{{Hockney} \&
  {Eastwood}}{1988}]{hockney88}
{Hockney} R.~W.,  {Eastwood} J.~W.,  1988, {Computer simulation using
  particles}

\bibitem[\protect\citeauthoryear{{Hopkins}}{{Hopkins}}{2013}]{hopkins13}
{Hopkins} P.~F.,  2013, \mn@doi [\mnras] {10.1093/mnras/sts210}, \href
  {http://adsabs.harvard.edu/abs/2013MNRAS.428.2840H} {428, 2840}

\bibitem[\protect\citeauthoryear{{Hopkins}}{{Hopkins}}{2015}]{hopkins15a}
{Hopkins} P.~F.,  2015, \mn@doi [\mnras] {10.1093/mnras/stv195}, \href
  {http://adsabs.harvard.edu/abs/2015MNRAS.450...53H} {450, 53}

\bibitem[\protect\citeauthoryear{{Hopkins} \& {Raives}}{{Hopkins} \&
  {Raives}}{2015}]{hopkins15b}
{Hopkins} P.~F.,  {Raives} M.~J.,  2015, preprint, \href
  {http://adsabs.harvard.edu/abs/2015arXiv150502783H} {} (\mn@eprint {arXiv}
  {1505.02783})

\bibitem[\protect\citeauthoryear{{Hu}, {Naab}, {Walch}, {Moster}  \&
  {Oser}}{{Hu} et~al.}{2014}]{hu14}
{Hu} C.-Y.,  {Naab} T.,  {Walch} S.,  {Moster} B.~P.,   {Oser} L.,  2014,
  \mn@doi [\mnras] {10.1093/mnras/stu1187}, \href
  {http://adsabs.harvard.edu/abs/2014MNRAS.443.1173H} {443, 1173}

\bibitem[\protect\citeauthoryear{{Jasche}, {Kitaura}  \& {Ensslin}}{{Jasche}
  et~al.}{2009}]{jasche09}
{Jasche} J.,  {Kitaura} F.~S.,   {Ensslin} T.~A.,  2009, preprint, \href
  {http://adsabs.harvard.edu/abs/2009arXiv0901.3043J} {} (\mn@eprint {arXiv}
  {0901.3043})

\bibitem[\protect\citeauthoryear{{Jin}, {Krokos}, {Rivi}, {Gheller}, {Dolag}
  \& {Reinecke}}{{Jin} et~al.}{2010}]{jin10}
{Jin} Z.,  {Krokos} M.,  {Rivi} M.,  {Gheller} C.,  {Dolag} K.,   {Reinecke}
  M.,  2010, preprint, \href
  {http://adsabs.harvard.edu/abs/2010arXiv1004.1302J} {} (\mn@eprint {arXiv}
  {1004.1302})

\bibitem[\protect\citeauthoryear{{Jing}}{{Jing}}{2005}]{jing05}
{Jing} Y.~P.,  2005, \mn@doi [\apj] {10.1086/427087}, \href
  {http://adsabs.harvard.edu/abs/2005ApJ...620..559J} {620, 559}

\bibitem[\protect\citeauthoryear{{Junk}, {Walch}, {Heitsch}, {Burkert},
  {Wetzstein}, {Schartmann}  \& {Price}}{{Junk} et~al.}{2010}]{junk10}
{Junk} V.,  {Walch} S.,  {Heitsch} F.,  {Burkert} A.,  {Wetzstein} M.,
  {Schartmann} M.,   {Price} D.,  2010, \mn@doi [\mnras]
  {10.1111/j.1365-2966.2010.17039.x}, \href
  {http://adsabs.harvard.edu/abs/2010MNRAS.407.1933J} {407, 1933}

\bibitem[\protect\citeauthoryear{{Komatsu} \& {Seljak}}{{Komatsu} \&
  {Seljak}}{2001}]{komatsu01}
{Komatsu} E.,  {Seljak} U.,  2001, \mn@doi [\mnras]
  {10.1046/j.1365-8711.2001.04838.x}, \href
  {http://adsabs.harvard.edu/abs/2001MNRAS.327.1353K} {327, 1353}

\bibitem[\protect\citeauthoryear{{Kravtsov} \& {Borgani}}{{Kravtsov} \&
  {Borgani}}{2012}]{kravtsov12}
{Kravtsov} A.~V.,  {Borgani} S.,  2012, \mn@doi [\araa]
  {10.1146/annurev-astro-081811-125502}, \href
  {http://adsabs.harvard.edu/abs/2012ARA%26A..50..353K} {50, 353}

\bibitem[\protect\citeauthoryear{{Landau} \& {Lifshitz}}{{Landau} \&
  {Lifshitz}}{1959}]{landau59}
{Landau} L.~D.,  {Lifshitz} E.~M.,  1959, {Fluid mechanics}

\bibitem[\protect\citeauthoryear{{Lucy}}{{Lucy}}{1977}]{lucy77}
{Lucy} L.~B.,  1977, \mn@doi [\aj] {10.1086/112164}, \href
  {http://adsabs.harvard.edu/abs/1977AJ.....82.1013L} {82, 1013}

\bibitem[\protect\citeauthoryear{{McNally}, {Lyra}  \& {Passy}}{{McNally}
  et~al.}{2012}]{mcnally12}
{McNally} C.~P.,  {Lyra} W.,   {Passy} J.-C.,  2012, \mn@doi [\apjs]
  {10.1088/0067-0049/201/2/18}, \href
  {http://adsabs.harvard.edu/abs/2012ApJS..201...18M} {201, 18}

\bibitem[\protect\citeauthoryear{{Monaghan}}{{Monaghan}}{1992}]{monaghan92}
{Monaghan} J.~J.,  1992, \mn@doi [\araa] {10.1146/annurev.aa.30.090192.002551},
  \href {http://adsabs.harvard.edu/abs/1992ARA%26A..30..543M} {30, 543}

\bibitem[\protect\citeauthoryear{{Monaghan}}{{Monaghan}}{1997}]{monaghan97}
{Monaghan} J.~J.,  1997, \mn@doi [Journal of Computational Physics]
  {10.1006/jcph.1997.5732}, \href
  {http://adsabs.harvard.edu/abs/1997JCoPh.136..298M} {136, 298}

\bibitem[\protect\citeauthoryear{{Monaghan}}{{Monaghan}}{2012}]{monaghan12}
{Monaghan} J.~J.,  2012, \mn@doi [Annual Review of Fluid Mechanics]
  {10.1146/annurev-fluid-120710-101220}, \href
  {http://adsabs.harvard.edu/abs/2012AnRFM..44..323M} {44, 323}

\bibitem[\protect\citeauthoryear{{Monaghan} \& {Gingold}}{{Monaghan} \&
  {Gingold}}{1983}]{monaghan83}
{Monaghan} J.~J.,  {Gingold} R.~A.,  1983, \mn@doi [Journal of Computational
  Physics] {10.1016/0021-9991(83)90036-0}, \href
  {http://adsabs.harvard.edu/abs/1983JCoPh..52..374M} {52, 374}

\bibitem[\protect\citeauthoryear{{Monaghan} \& {Lattanzio}}{{Monaghan} \&
  {Lattanzio}}{1985}]{monaghan85}
{Monaghan} J.~J.,  {Lattanzio} J.~C.,  1985, \aap, \href
  {http://adsabs.harvard.edu/abs/1985A%26A...149..135M} {149, 135}

\bibitem[\protect\citeauthoryear{{Morris} \& {Monaghan}}{{Morris} \&
  {Monaghan}}{1997}]{morris97}
{Morris} J.~P.,  {Monaghan} J.~J.,  1997, \mn@doi [Journal of Computational
  Physics] {10.1006/jcph.1997.5690}, \href
  {http://adsabs.harvard.edu/abs/1997JCoPh.136...41M} {136, 41}

\bibitem[\protect\citeauthoryear{{Murante}, {Borgani}, {Brunino}  \&
  {Cha}}{{Murante} et~al.}{2011}]{murante11}
{Murante} G.,  {Borgani} S.,  {Brunino} R.,   {Cha} S.-H.,  2011, \mn@doi
  [\mnras] {10.1111/j.1365-2966.2011.19021.x}, \href
  {http://adsabs.harvard.edu/abs/2011MNRAS.417..136M} {417, 136}

\bibitem[\protect\citeauthoryear{{Navarro}, {Frenk}  \& {White}}{{Navarro}
  et~al.}{1997}]{navarro97}
{Navarro} J.~F.,  {Frenk} C.~S.,   {White} S.~D.~M.,  1997, \apj, \href
  {http://adsabs.harvard.edu/abs/1997ApJ...490..493N} {490, 493}

\bibitem[\protect\citeauthoryear{{Pakmor}}{{Pakmor}}{2010}]{pakmor10}
{Pakmor} R.,  2010, {Progenitor systems of Type Ia Supernovae: mergers of white
  dwarfs and constraints on hydrogen-accreting white dwarfs}

\bibitem[\protect\citeauthoryear{{Pakmor}, {Edelmann}, {R{\"o}pke}  \&
  {Hillebrandt}}{{Pakmor} et~al.}{2012}]{pakmor12}
{Pakmor} R.,  {Edelmann} P.,  {R{\"o}pke} F.~K.,   {Hillebrandt} W.,  2012,
  \mn@doi [\mnras] {10.1111/j.1365-2966.2012.21383.x}, \href
  {http://adsabs.harvard.edu/abs/2012MNRAS.424.2222P} {424, 2222}

\bibitem[\protect\citeauthoryear{{Planelles} \& {Quilis}}{{Planelles} \&
  {Quilis}}{2009}]{planelles09}
{Planelles} S.,  {Quilis} V.,  2009, \mn@doi [\mnras]
  {10.1111/j.1365-2966.2009.15290.x}, \href
  {http://adsabs.harvard.edu/abs/2009MNRAS.399..410P} {399, 410}

\bibitem[\protect\citeauthoryear{{Power}, {Read}  \& {Hobbs}}{{Power}
  et~al.}{2014}]{power14}
{Power} C.,  {Read} J.~I.,   {Hobbs} A.,  2014, \mn@doi [\mnras]
  {10.1093/mnras/stu418}, \href
  {http://adsabs.harvard.edu/abs/2014MNRAS.440.3243P} {440, 3243}

\bibitem[\protect\citeauthoryear{{Price}}{{Price}}{2007}]{price07}
{Price} D.~J.,  2007, \mn@doi [\pasa] {10.1071/AS07022}, \href
  {http://adsabs.harvard.edu/abs/2007PASA...24..159P} {24, 159}

\bibitem[\protect\citeauthoryear{{Price}}{{Price}}{2008}]{price08}
{Price} D.~J.,  2008, \mn@doi [Journal of Computational Physics]
  {10.1016/j.jcp.2008.08.011}, \href
  {http://adsabs.harvard.edu/abs/2008JCoPh.22710040P} {227, 10040}

\bibitem[\protect\citeauthoryear{{Price}}{{Price}}{2012a}]{price12a}
{Price} D.~J.,  2012a, \mn@doi [Journal of Computational Physics]
  {10.1016/j.jcp.2010.12.011}, \href
  {http://adsabs.harvard.edu/abs/2012JCoPh.231..759P} {231, 759}

\bibitem[\protect\citeauthoryear{{Price}}{{Price}}{2012b}]{price12b}
{Price} D.~J.,  2012b, \mn@doi [\mnras] {10.1111/j.1745-3933.2011.01187.x},
  \href {http://adsabs.harvard.edu/abs/2012MNRAS.420L..33P} {420, L33}

\bibitem[\protect\citeauthoryear{{Puri} \& {Ramachandran}}{{Puri} \&
  {Ramachandran}}{2014}]{puri13}
{Puri} K.,  {Ramachandran} P.,  2014, \mn@doi [Journal of Computational
  Physics] {10.1016/j.jcp.2014.03.055}, \href
  {http://adsabs.harvard.edu/abs/2014JCoPh.270..432P} {270, 432}

\bibitem[\protect\citeauthoryear{{Quilis}}{{Quilis}}{2004}]{quilis04}
{Quilis} V.,  2004, \mn@doi [\mnras] {10.1111/j.1365-2966.2004.08040.x}, \href
  {http://adsabs.harvard.edu/abs/2004MNRAS.352.1426Q} {352, 1426}

\bibitem[\protect\citeauthoryear{{Rasia} et~al.,}{{Rasia}
  et~al.}{2015}]{rasia15}
{Rasia} E.,  et~al., 2015, preprint, \href
  {http://adsabs.harvard.edu/abs/2015arXiv150904247R} {} (\mn@eprint {arXiv}
  {1509.04247})

\bibitem[\protect\citeauthoryear{{Read} \& {Hayfield}}{{Read} \&
  {Hayfield}}{2012}]{read12}
{Read} J.~I.,  {Hayfield} T.,  2012, \mn@doi [\mnras]
  {10.1111/j.1365-2966.2012.20819.x}, \href
  {http://adsabs.harvard.edu/abs/2012MNRAS.422.3037R} {422, 3037}

\bibitem[\protect\citeauthoryear{{Read}, {Hayfield}  \& {Agertz}}{{Read}
  et~al.}{2010}]{read10}
{Read} J.~I.,  {Hayfield} T.,   {Agertz} O.,  2010, \mn@doi [\mnras]
  {10.1111/j.1365-2966.2010.16577.x}, \href
  {http://adsabs.harvard.edu/abs/2010MNRAS.405.1513R} {405, 1513}

\bibitem[\protect\citeauthoryear{{Ritchie} \& {Thomas}}{{Ritchie} \&
  {Thomas}}{2001}]{ritchie01}
{Ritchie} B.~W.,  {Thomas} P.~A.,  2001, \mn@doi [\mnras]
  {10.1046/j.1365-8711.2001.04268.x}, \href
  {http://adsabs.harvard.edu/abs/2001MNRAS.323..743R} {323, 743}

\bibitem[\protect\citeauthoryear{{Rosswog}}{{Rosswog}}{2009}]{rosswog09}
{Rosswog} S.,  2009, \mn@doi [\nar] {10.1016/j.newar.2009.08.007}, \href
  {http://adsabs.harvard.edu/abs/2009NewAR..53...78R} {53, 78}

\bibitem[\protect\citeauthoryear{{Rosswog}}{{Rosswog}}{2015}]{rosswog14}
{Rosswog} S.,  2015, \mn@doi [\mnras] {10.1093/mnras/stv225}, \href
  {http://adsabs.harvard.edu/abs/2015MNRAS.448.3628R} {448, 3628}

\bibitem[\protect\citeauthoryear{{Saitoh} \& {Makino}}{{Saitoh} \&
  {Makino}}{2009}]{saitoh09}
{Saitoh} T.~R.,  {Makino} J.,  2009, \mn@doi [\apjl]
  {10.1088/0004-637X/697/2/L99}, \href
  {http://adsabs.harvard.edu/abs/2009ApJ...697L..99S} {697, L99}

\bibitem[\protect\citeauthoryear{{Saitoh} \& {Makino}}{{Saitoh} \&
  {Makino}}{2013}]{saitoh13}
{Saitoh} T.~R.,  {Makino} J.,  2013, \mn@doi [\apj]
  {10.1088/0004-637X/768/1/44}, \href
  {http://adsabs.harvard.edu/abs/2013ApJ...768...44S} {768, 44}

\bibitem[\protect\citeauthoryear{{Scannapieco} et~al.,}{{Scannapieco}
  et~al.}{2012}]{scannapieco12}
{Scannapieco} C.,  et~al., 2012, \mn@doi [\mnras]
  {10.1111/j.1365-2966.2012.20993.x}, \href
  {http://adsabs.harvard.edu/abs/2012MNRAS.423.1726S} {423, 1726}

\bibitem[\protect\citeauthoryear{{Schaye} et~al.,}{{Schaye}
  et~al.}{2015}]{schaye15}
{Schaye} J.,  et~al., 2015, \mn@doi [\mnras] {10.1093/mnras/stu2058}, \href
  {http://adsabs.harvard.edu/abs/2015MNRAS.446..521S} {446, 521}

\bibitem[\protect\citeauthoryear{{Schuessler} \& {Schmitt}}{{Schuessler} \&
  {Schmitt}}{1981}]{schuessler81}
{Schuessler} I.,  {Schmitt} D.,  1981, \aap, \href
  {http://adsabs.harvard.edu/abs/1981A%26A....97..373S} {97, 373}

\bibitem[\protect\citeauthoryear{{Sedov}}{{Sedov}}{1959}]{sedov59}
{Sedov} L.~I.,  1959, {Similarity and Dimensional Methods in Mechanics}

\bibitem[\protect\citeauthoryear{{Sembolini} et~al.,}{{Sembolini}
  et~al.}{2015}]{sembolini15}
{Sembolini} F.,  et~al., 2015, preprint, \href
  {http://adsabs.harvard.edu/abs/2015arXiv150306065S} {} (\mn@eprint {arXiv}
  {1503.06065})

\bibitem[\protect\citeauthoryear{{Shen}, {Wadsley}  \& {Stinson}}{{Shen}
  et~al.}{2010}]{shen10}
{Shen} S.,  {Wadsley} J.,   {Stinson} G.,  2010, \mn@doi [\mnras]
  {10.1111/j.1365-2966.2010.17047.x}, \href
  {http://adsabs.harvard.edu/abs/2010MNRAS.407.1581S} {407, 1581}

\bibitem[\protect\citeauthoryear{{Sod}}{{Sod}}{1978}]{sod78}
{Sod} G.~A.,  1978, \mn@doi [Journal of Computational Physics]
  {10.1016/0021-9991(78)90023-2}, \href
  {http://adsabs.harvard.edu/abs/1978JCoPh..27....1S} {27, 1}

\bibitem[\protect\citeauthoryear{{Springel}}{{Springel}}{2005}]{springel05}
{Springel} V.,  2005, \mn@doi [\mnras] {10.1111/j.1365-2966.2005.09655.x},
  \href {http://adsabs.harvard.edu/abs/2005MNRAS.364.1105S} {364, 1105}

\bibitem[\protect\citeauthoryear{{Springel}}{{Springel}}{2010a}]{springel10a}
{Springel} V.,  2010a, \mn@doi [\araa] {10.1146/annurev-astro-081309-130914},
  \href {http://adsabs.harvard.edu/abs/2010ARA%26A..48..391S} {48, 391}

\bibitem[\protect\citeauthoryear{{Springel}}{{Springel}}{2010b}]{springel10b}
{Springel} V.,  2010b, \mn@doi [\mnras] {10.1111/j.1365-2966.2009.15715.x},
  \href {http://adsabs.harvard.edu/abs/2010MNRAS.401..791S} {401, 791}

\bibitem[\protect\citeauthoryear{{Springel} \& {Hernquist}}{{Springel} \&
  {Hernquist}}{2002}]{springel02}
{Springel} V.,  {Hernquist} L.,  2002, \mn@doi [\mnras]
  {10.1046/j.1365-8711.2002.05445.x}, \href
  {http://adsabs.harvard.edu/abs/2002MNRAS.333..649S} {333, 649}

\bibitem[\protect\citeauthoryear{{Springel} \& {Hernquist}}{{Springel} \&
  {Hernquist}}{2003}]{springel03}
{Springel} V.,  {Hernquist} L.,  2003, \mn@doi [\mnras]
  {10.1046/j.1365-8711.2003.06206.x}, \href
  {http://adsabs.harvard.edu/abs/2003MNRAS.339..289S} {339, 289}

\bibitem[\protect\citeauthoryear{{Springel}, {Yoshida}  \& {White}}{{Springel}
  et~al.}{2001}]{springel01}
{Springel} V.,  {Yoshida} N.,   {White} S.~D.~M.,  2001, \mn@doi [\na]
  {10.1016/S1384-1076(01)00042-2}, \href
  {http://adsabs.harvard.edu/abs/2001NewA....6...79S} {6, 79}

\bibitem[\protect\citeauthoryear{{Steinmetz} \& {Mueller}}{{Steinmetz} \&
  {Mueller}}{1993}]{steinmetz93}
{Steinmetz} M.,  {Mueller} E.,  1993, \aap, \href
  {http://esoads.eso.org/abs/1993A%26A...268..391S} {268, 391}

\bibitem[\protect\citeauthoryear{{Stone}, {Gardiner}, {Teuben}, {Hawley}  \&
  {Simon}}{{Stone} et~al.}{2008}]{stone08}
{Stone} J.~M.,  {Gardiner} T.~A.,  {Teuben} P.,  {Hawley} J.~F.,   {Simon}
  J.~B.,  2008, \mn@doi [\apjs] {10.1086/588755}, \href
  {http://adsabs.harvard.edu/abs/2008ApJS..178..137S} {178, 137}

\bibitem[\protect\citeauthoryear{{Trac} \& {Pen}}{{Trac} \&
  {Pen}}{2004}]{trac04}
{Trac} H.,  {Pen} U.-L.,  2004, \mn@doi [\na] {10.1016/j.newast.2004.02.002},
  \href {http://adsabs.harvard.edu/abs/2004NewA....9..443T} {9, 443}

\bibitem[\protect\citeauthoryear{{Tricco} \& {Price}}{{Tricco} \&
  {Price}}{2013}]{tricco13}
{Tricco} T.,  {Price} D.,  2013, preprint, \href
  {http://adsabs.harvard.edu/abs/2013arXiv1310.4260T} {} (\mn@eprint {arXiv}
  {1310.4260})

\bibitem[\protect\citeauthoryear{{Valcke}, {de Rijcke}, {R{\"o}diger}  \&
  {Dejonghe}}{{Valcke} et~al.}{2010}]{valcke10}
{Valcke} S.,  {de Rijcke} S.,  {R{\"o}diger} E.,   {Dejonghe} H.,  2010,
  \mn@doi [\mnras] {10.1111/j.1365-2966.2010.17127.x}, \href
  {http://adsabs.harvard.edu/abs/2010MNRAS.408...71V} {408, 71}

\bibitem[\protect\citeauthoryear{{Valdarnini}}{{Valdarnini}}{2012}]{valdarnini12}
{Valdarnini} R.,  2012, \mn@doi [\aap] {10.1051/0004-6361/201219715}, \href
  {http://adsabs.harvard.edu/abs/2012A%26A...546A..45V} {546, A45}

\bibitem[\protect\citeauthoryear{{Vazza}}{{Vazza}}{2011}]{vazza11}
{Vazza} F.,  2011, \mn@doi [\mnras] {10.1111/j.1365-2966.2010.17455.x}, \href
  {http://adsabs.harvard.edu/abs/2011MNRAS.410..461V} {410, 461}

\bibitem[\protect\citeauthoryear{{Viola}, {Monaco}, {Borgani}, {Murante}  \&
  {Tornatore}}{{Viola} et~al.}{2008}]{viola08}
{Viola} M.,  {Monaco} P.,  {Borgani} S.,  {Murante} G.,   {Tornatore} L.,
  2008, \mn@doi [\mnras] {10.1111/j.1365-2966.2007.12598.x}, \href
  {http://adsabs.harvard.edu/abs/2008MNRAS.383..777V} {383, 777}

\bibitem[\protect\citeauthoryear{{Wadsley}, {Veeravalli}  \&
  {Couchman}}{{Wadsley} et~al.}{2008}]{wadsley08}
{Wadsley} J.~W.,  {Veeravalli} G.,   {Couchman} H.~M.~P.,  2008, \mn@doi
  [\mnras] {10.1111/j.1365-2966.2008.13260.x}, \href
  {http://adsabs.harvard.edu/abs/2008MNRAS.387..427W} {387, 427}

\bibitem[\protect\citeauthoryear{{White}}{{White}}{1996}]{white96}
{White} S.~D.~M.,  1996, in {Schaeffer} R.,  {Silk} J.,  {Spiro} M.,
  {Zinn-Justin} J.,  eds, Cosmology and Large Scale Structure. p.~349

\bibitem[\protect\citeauthoryear{{Zel'dovich}}{{Zel'dovich}}{1970}]{zeldovich70}
{Zel'dovich} Y.~B.,  1970, \aap, \href
  {http://adsabs.harvard.edu/abs/1970A%26A.....5...84Z} {5, 84}

\makeatother
\end{thebibliography}

\bsp

\label{lastpage}

\end{document}